# Magnetostatic interaction in oriented assembly of elongated nanoparticles

N. A. Usov[1,2]

[1]*National University of Science and Technology «MISiS», 119049, Moscow, Russia*

[2]*Pushkov Institute of Terrestrial Magnetism, Ionosphere and Radio Wave Propagation, Russian Academy of Sciences, IZMIRAN, 108480, Troitsk, Moscow, Russia*

**Abstract.** The elements of the magneto- dipole (MD) interaction matrix are calculated for a pair of oriented spheroidal magnetic nanoparticles with a semiaxes ratio $a/b$ = 1.25, 1.5, 2.0 and 3.0 as a function of the distance between the particle centers. It is shown that the spherical approximation for MD interaction matrix is incorrect already at aspect ratios $a/b > 1.25$, if the distance $R$ between the particle centers is of the order of particle sizes. However, for moderate particle aspect ratios, $a/b < 1.5$, the first order correction to spherical approximation with respect to a small parameter $(a^2 - b^2)/R^2$ is shown to be in a good agreement with numerical data. Using exact MD interaction matrix, the quasi-static hysteresis loops of dilute assemblies of oriented clusters of spheroidal nanoparticles with different filling densities are calculated, depending on the direction of the external magnetic field with respect to the particle orientation axis.

# 1. Introduction

Assemblies of magnetic nanoparticles are promising for various fields of modern nanotechnology [1-3], as well as biomedicine [4-6]. However, it is important to control thoroughly the geometric and magnetic characteristics of nanoparticles for successful development of these applications. In theoretical modeling of the properties of magnetic nanoparticle assemblies [7-16] the results of interest for technical applications are obtained, as a rule, on the assumption that the particles are homogeneous in composition, single-crystal and have a narrow size distribution. Unfortunately, the existing methods for creating assemblies of magnetic nanoparticles [3,5, 17-20] are still not sufficiently perfect. Indeed, nanoparticles obtained by various techniques are usually polycrystalline, irregular in shape, with a wide size distribution [21-23]. It is worth noting also that in dense nanoparticle assemblies the mutual magneto- dipole (MD) interaction has a significant influence on the assembly properties [24-26]. Meanwhile, the effect of the MD interaction is usually theoretically modeled [7-16] assuming the outer shape of the particles to be close to quasispherical. However, this supposition is often far from reality [19,20, 27-30]. The outer shape of an elongated nanoparticle seems close to that of an ellipsoid, or spheroid, with a certain semiaxes ratio $a/b > 1$. Unfortunately, explicit formulas for the elements of the MD interaction matrix are available only for spherical nanoparticles. For particles of ellipsoidal or spheroidal shape there are only complicated expressions represented by multidimensional integrals over volumes or over particle surfaces [31–32].

To quantitatively estimate the effect of the MD interaction on the properties of a dense oriented assembly of spheroidal nanoparticles, in this work the tabular values for the elements of MD interaction matrix are obtained numerically for a pair of oriented spheroidal nanoparticles with different aspect ratios $a/b$ = 1.25, 1.5, 2.0 and 3.0 depending on the distance between the particle centers. These data made it possible to calculate the quasi-static hysteresis loops of oriented assemblies of clusters of spheroidal nanoparticles with different aspect ratios for various particle filling density $\eta$ within the clusters.

It is shown that the spherical approximation for the elements of the MD interaction matrix is incorrect already at aspect ratio $a/b > 1.25$, if the distance $R$ between the particle centers is of the order of particle sizes. In addition to the numerical results, the first order correction to the spherical approximation with respect to a small parameter $(a^2 - b^2)/R^2$ is obtained analytically. This approximation is shown to be in a good agreement with the numerical data for moderate particle aspect ratios, $a/b < 1.5$.

## 2. Interaction matrix of oriented assembly

In a theoretical study of the magnetic properties of a dense assembly of elongated single-domain nanoparticles, the greatest difficulty is caused by calculating the MD interaction energy of nanoparticles depending on the distance between particles and their mutual orientation. It is known that the MD interaction energy of a pair of spherical particles coincides with that of point magnetic dipoles [31]

$$W_m = M_s^2 V^2 \frac{\vec{\alpha}_1 \vec{\alpha}_2 - 3(\vec{n}\vec{\alpha}_1)(\vec{n}\vec{\alpha}_2)}{R^3} = \sum_{i,j} A_{ij} \alpha_i^{(1)} \alpha_j^{(2)}, \qquad (i,j = x, y, z), \qquad (1)$$

where $M_s$ is the saturation magnetization, $V$ is the particle volume, $\vec{\alpha}_1$ and $\vec{\alpha}_2$ are the unit magnetization vectors of the first and second single-domain nanoparticles, respectively, $\vec{n}$ is the unit vector connecting nanoparticle centers, and $A_{ij}$ are the elements of the MD interaction matrix.

Without loss of generality, one can assume that the nanoparticle centers are located at the points (0,0,0) and $\vec{R} = (X_0, 0, Z_0)$ of the Cartesian coordinates. Then the unit vector $\vec{n}$ is

$$\vec{n} = \left(\frac{X_0}{R}, 0, \frac{Z_0}{R}\right); \qquad R = \sqrt{X_0^2 + Z_0^2}. \qquad (2)$$

From Eqs. (1), (2) one can obtain the nonzero elements of the MD interaction matrix $A_{ij}$ in the considered case as follows

$$A_{xx} = \frac{M_s^2 V^2}{R^3}(1 - 3n_x^2); \quad A_{yy} = \frac{M_s^2 V^2}{R^3}; \quad A_{zz} = \frac{M_s^2 V^2}{R^3}(1 - 3n_z^2); \quad A_{xz} = A_{zx} = -3\frac{M_s^2 V^2}{R^3} n_x n_z. \qquad (3)$$

All other off-diagonal elements of this matrix are identically zero, $A_{xy} = A_{yx} = A_{yz} = A_{zy} = 0$.

However, even for a simplest case of single-domain nanoparticles of spheroidal shape there is no explicit expression for the elements of the corresponding MD interaction matrix. This energy can be written, for example, in the form of a six-dimensional integral over the particle volumes, [32]

$$W_m = M_s^2 \int_V (\vec{\alpha}_1 \vec{\nabla}_1) \int_V (\vec{\alpha}_2 \vec{\nabla}_2) \frac{dv_1 dv_2}{|\vec{r}_1 - \vec{r}_2|} = \sum_{i,j} B_{ij} \alpha_i^{(1)} \alpha_j^{(2)}, \qquad (i,j = x, y, z), \qquad (4)$$

where $B_{ij}$ are the elements of the MD interaction matrix for spheroidal nanoparticles.

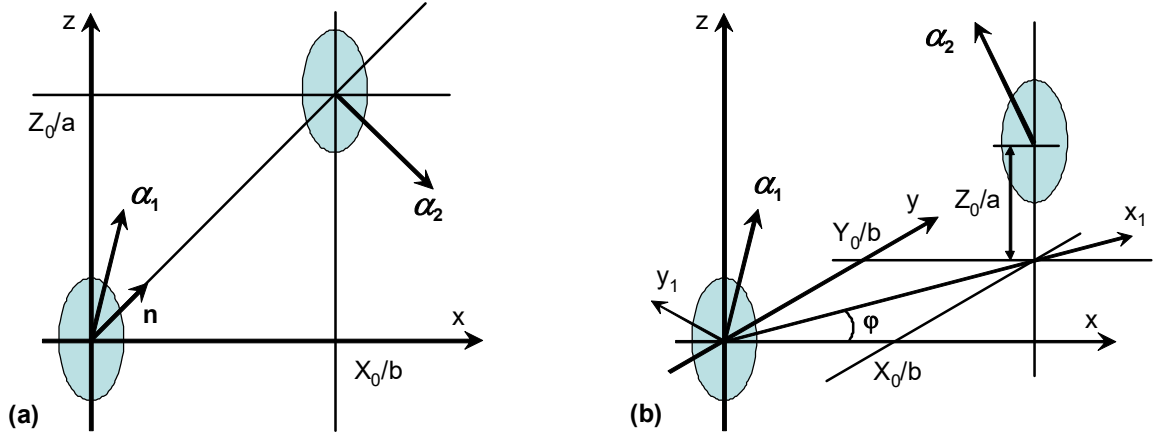

Fig. 1. a) Location of a pair of oriented spheroidal magnetic nanoparticles on the *XZ* plane with centers at points (0,0,0) and ($X_0$,0,$Z_0$), respectively. b) The general case of the arrangement of oriented spheroidal nanoparticles with centers at the points (0,0,0) and ($X_0$,$Y_0$,$Z_0$).

For a pair of spheroidal nanoparticles with aspect ratio $a/b > 1$ and volume $V = 4\pi a b^2/3$, oriented along the *Z* axis, the expressions for nonzero matrix elements of MD interaction matrix, convenient for numerical integration, can be obtained as follows. Consider first an auxiliary case, when the center of the first spheroidal particle is at the origin of Cartesian coordinates, whereas the center of the second particle is determined by the displacement vector $\boldsymbol{R} = (X_0, 0, Z_0)$ (see Fig. 1a).

Based on Eq. (4), for the matrix element $B_{xx}$ one obtains

$$B_{xx}(X_0, Z_0) = M_s^2 \int_V dv_1 \frac{\partial}{\partial x_1} \int_V dv_2 \frac{\partial}{\partial x_2} \frac{1}{\sqrt{(x_1 - x_2 - X_0)^2 + (y_1 - y_2)^2 + (z_1 - z_2 - Z_0)^2}}. \qquad (5)$$

Let us introduce dimensionless variables in this integral setting $\bar{x}_i = x_i/b$, $\bar{y}_i = y_i/b$, $\bar{z}_i = z_i/a$, $i = 1,2$. Integrating over $\bar{x}_1$ and $\bar{x}_2$ variables by parts, one obtains the normalized value of the element $B_{xx}$ in the form

$$\frac{B_{xx}}{M_s^2 a^2 b} = \int dS_1 \int dS_2 \left\{ \frac{1}{\sqrt{(t_1 - t_2 - X_0/b)^2 + q_{xx}^2}} - \frac{1}{\sqrt{(t_1 + t_2 - X_0/b)^2 + q_{xx}^2}} - \right.$$

$$\left.\frac{1}{\sqrt{(t_1 + t_2 + X_0/b)^2 + q_{xx}^2}} + \frac{1}{\sqrt{(t_1 - t_2 + X_0/b)^2 + q_{xx}^2}}\right\}. \tag{6}$$

Here we denote $q_{xx}^2 = (\bar{y}_1 - \bar{y}_2)^2 + (a/b)^2(\bar{z}_1 - \bar{z}_2 - Z_0/a)^2$, and $\bar{y}_1 = \rho_1 \cos\varphi_1$, $\bar{z}_1 = \rho_1 \sin\varphi_1$, $\bar{y}_2 = \rho_2 \cos\varphi_2$, $\bar{z}_2 = \rho_2 \sin\varphi_2$, $t_1 = \sqrt{1 - \rho_1^2}$, $t_2 = \sqrt{1 - \rho_2^2}$. In Eq. (6) the integration is performed over the area of circles of unit radius, so that $dS_1 = \rho_1 d\rho_1 d\varphi_1$, $dS_2 = \rho_2 d\rho_2 d\varphi_2$.

Similar expressions can be obtained for other nonzero elements of the interaction matrix, namely

$$\frac{B_{yy}}{M_S^2 a^2 b} = 2\int dS_1 \int dS_2 \left\{ \frac{1}{\sqrt{(t_1 - t_2)^2 + q_{yy}^2}} - \frac{1}{\sqrt{(t_1 + t_2)^2 + q_{yy}^2}} \right\}, \tag{7}$$

where $q_{yy}^2 = (\bar{x}_1 - \bar{x}_2 - X_0/b)^2 + (a/b)^2(\bar{y}_1 - \bar{y}_2 - Z_0/a)^2$,

$$\frac{B_{zz}}{M_S^2 a^2 b} = \left(\frac{b}{a}\right)^2 \int dS_1 \int dS_2 \left\{ \frac{1}{\sqrt{(a/b)^2(t_1 - t_2 - Z_0/a)^2 + q_{zz}^2}} - \frac{1}{\sqrt{(a/b)^2(t_1 + t_2 + Z_0/a)^2 + q_{zz}^2}} - \right.$$

$$\left. \frac{1}{\sqrt{(a/b)^2(t_1 + t_2 - Z_0/a)^2 + q_{zz}^2}} + \frac{1}{\sqrt{(a/b)^2(t_1 - t_2 + Z_0/a)^2 + q_{zz}^2}} \right\}, \tag{8}$$

where $q_{zz}^2 = (\bar{x}_1 - \bar{x}_2 - X_0/b)^2 + (\bar{y}_1 - \bar{y}_2)^2$, and finally

$$\frac{B_{xz}}{M_S^2 a^2 b} = \frac{b}{a} \int dS_1 \int dS_2 \left\{ \frac{1}{\sqrt{(t_1 - \bar{x}_2 - X_0/b)^2 + q_{xz}^2 + (a/b)^2(\bar{x}_1 - t_2 - Z_0/a)^2}} - \right.$$

$$\frac{1}{\sqrt{(t_1 + \bar{x}_2 + X_0/b)^2 + q_{xz}^2 + (a/b)^2(\bar{x}_1 - t_2 - Z_0/a)^2}} - \frac{1}{\sqrt{(t_1 - \bar{x}_2 - X_0/b)^2 + q_{xz}^2 + (a/b)^2(\bar{x}_1 + t_2 - Z_0/a)^2}} +$$

$$\left. \frac{1}{\sqrt{(t_1 + \bar{x}_2 + X_0/b)^2 + q_{xz}^2 + (a/b)^2(\bar{x}_1 + t_2 - Z_0/a)^2}} \right\}, \tag{9}$$

where $q_{xz}^2 = (\bar{y}_1 - \bar{y}_2)^2$.

Since the volumes of spheroidal nanoparticles do not intersect by condition, the integrals (6) - (9) have no singularities in the denominators of the integrands and can be calculated using standard numerical integration programs. It can be shown that the element $B_{zx} = B_{xz}$, whereas the remaining off-diagonal elements of the interaction matrix are equal to zero due to the symmetry of the problem, $B_{xy} = B_{yx} = B_{yz} = B_{zy} = 0$.

From the structure of integrals (6) - (9) one can deduce that the diagonal elements of the interaction matrix are even functions of the variables $X_0$ and $Z_0$

$$B_{xx}(-X_0, Z_0) = B_{xx}(X_0, Z_0), \qquad B_{xx}(X_0, -Z_0) = B_{xx}(X_0, Z_0), \qquad (10)$$

and similarly for the elements $B_{yy}$ and $B_{zz}$. At the same time, the off-diagonal element $B_{xz}$ is an odd function of these variables

$$B_{xz}(-X_0, Z_0) = -B_{xz}(X_0, Z_0), \qquad B_{xz}(X_0, -Z_0) = -B_{xz}(X_0, Z_0). \qquad (11)$$

Therefore, it suffices to calculate the elements $B_{xx}$, $B_{yy}$, $B_{zz}$ and $B_{xz}$ only for nonnegative displacements $X_0 \geq 0$ and $Z_0 \geq 0$. It is worth noting that, as follows from Eqs. (10), (11), the matrix elements $B_{ij}$ have the same symmetry properties as the matrix elements $A_{ij}$ for spherical nanoparticles, Eq. (3).

In this work detailed calculations have been carried out to obtain Tables S1 – S4 of nonzero elements of the interaction matrices $B_{xx}$, $B_{yy}$, $B_{zz}$ and $B_{xz}$ as the functions of the reduced variables $X_0/b$, $Z_0/a$ with a step of 0.5 for spheroidal nanoparticles with aspect ratios $a/b = 1.25, 1.5, 2.0,$ and $3.0$, respectively. The resulting Tables S1 – S4 are given in the Supplementary Information file.

To compare the elements $B_{ij}$ with the corresponding elements of the interaction matrix $A_{ij}$ for spherical nanoparticles of equal volume the latter can be conveniently written in the form

$$\frac{A_{xx}}{M_s^2 a^2 b} = \frac{(4\pi/3)^2}{\left((X_0/b)^2 + (a/b)^2(Z_0/a)^2\right)^{3/2}} \left(1 - 3\frac{(X_0/b)^2}{(X_0/b)^2 + (a/b)^2(Z_0/a)^2}\right), \qquad (12)$$

and similarly for the matrix elements $A_{yy}$, $A_{zz}$ and $A_{xz}$. These formulas can be considered as a spherical approximation for MD interaction matrix of spheroidal nanoparticles.

The Eqs. (6) - (9) for the elements of the interaction matrix $B_{xx}$, $B_{yy}$, $B_{zz}$ and $B_{xz}$ were obtained under the assumption that the centers of both spheroidal nanoparticles lie in a given *XZ* plane. To carry out three-dimensional micromagnetic calculations it is necessary to generalize these relations to the case when the center of the second nanoparticle is located at an arbitrary point $\vec{R} = (X_0, Y_0, Z_0)$ of the Cartesian coordinates, as shown in Fig. 1b. To determine the elements of the interaction matrix in this case, one can introduce a coordinate system $(X_1,Y_1,Z)$, rotated around the Z axis, so that the plane $(X_1,Z)$ passes through the center of the second particle. As seen in Fig. 1b, the angle of rotation of this coordinate system is determined by the equations

$$\sin\varphi = \frac{Y_0}{\sqrt{X_0^2 + Y_0^2}}, \qquad \cos\varphi = \frac{X_0}{\sqrt{X_0^2 + Y_0^2}}. \tag{13}$$

In the coordinate system $((X_1,Y_1,Z)$ the coordinates of the center of the second particle are $X_1 = \sqrt{X_0^2 + Y_0^2}$, $Y_1 = 0$. Therefore, the matrix elements of the auxiliary interaction matrix, $\tilde{B}_{x_1x_1}$, $\tilde{B}_{y_1y_1}$, $\tilde{B}_{zzz}$ and $\tilde{B}_{x_1z}$ can be calculated by formulas (6) - (9) for the corresponding displacements $X_1/b$, $Z_0/a$. Then the elements of the interaction matrix in the original coordinate system $(X,Y,Z)$ can be obtained by the formulas

$$B_{xx} = \tilde{B}_{x_1x_1}\cos^2\varphi + \tilde{B}_{y_1y_1}\sin^2\varphi, \qquad B_{yy} = \tilde{B}_{x_1x_1}\sin^2\varphi + \tilde{B}_{y_1y_1}\cos^2\varphi, \qquad B_{zz} = \tilde{B}_{zz}, \tag{14}$$

$$B_{xy} = B_{yx} = (\tilde{B}_{x_1x_1} - \tilde{B}_{y_1y_1})\sin\varphi\cos\varphi, \qquad B_{xz} = B_{zx} = \tilde{B}_{x_1z}\cos\varphi, \qquad B_{yz} = B_{zy} = \tilde{B}_{x_1z}\sin\varphi.$$

Taking into account Eqs. (13), (14), the numerical data presented in Tables S1 – S4 can be used to calculate the elements of the MD interaction matrix of an arbitrary pair of nanoparticles of an oriented assembly of spheroidal nanoparticles with a given aspect ratio $a/b$ = 1.25, 1.5, 2.0 and 3.0, respectively.

For the same arrangement of nanoparticles it is useful to compare the elements of the interaction matrix $B_{ij}$ of oriented spheroidal nanoparticles, Eqs. (6) – (9), with the matrix elements $A_{ij}$, Eq. (12), for spherical magnetic nanoparticles of equal volume. Fig. 2 shows a comparison of interaction matrix elements $B_{ij}$ and $A_{ij}$ for spheroidal nanoparticles with aspect ratio $a/b$ = 1.5 and spherical nanoparticles of equal volume, respectively, normalized to the value $M_s^2 a^2 b$. Dots in Fig. 2 show numerical results, solid lines demonstrate spherical approximation, Eq. (12). According to Fig. 2, the greatest difference between these matrix elements is observed for the domain of the reduced coordinates $0 \leq X_0/b \leq 3.0$ and $0 \leq Z_0/a \leq 2.0$, comparable with the reduced particle sizes. Outside this domain the spherical approximation works well for all matrix elements. However, for $X_0/b$, $Z_0/a \geq 4.0$ all matrix elements greatly reduce and become close to zero.

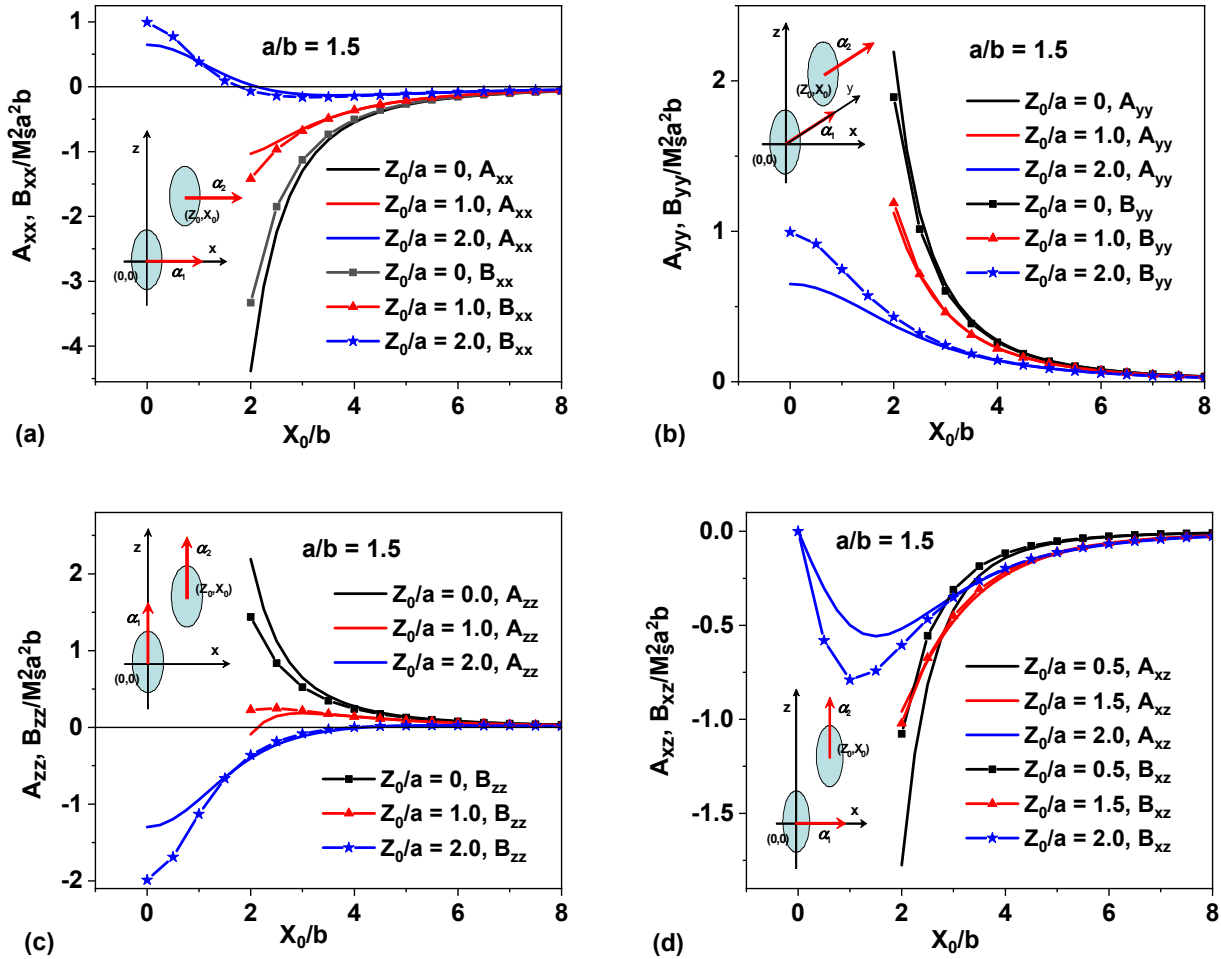

Fig. 2. Comparison of reduced MD interaction matrix elements for spheroidal nanoparticles with aspect ratio $a/b = 1.5$ and spherical nanoparticles of equal volume, respectively, for different values of the reduced coordinate $Z_0/a$. The inserts show the arrangement of nanoparticles on the $XZ$ plane and the corresponding orientation of their unit magnetization vectors.

According to Fig. 2, for the case $a/b = 1.5$ the behavior of the dependences $B_{ij}$ and $A_{ij}$ on the transverse reduced coordinate $X_0/b$ is qualitatively similar, although the quantitative difference between the corresponding matrix elements is noticeable at small values of $X_0/b$, $Z_0/a$. But for the case $a/b = 3.0$ shown in Fig. 3, the differences between the matrix elements $B_{ij}$ and $A_{ij}$ become very significant, so that the spherical approximation is certainly invalid.

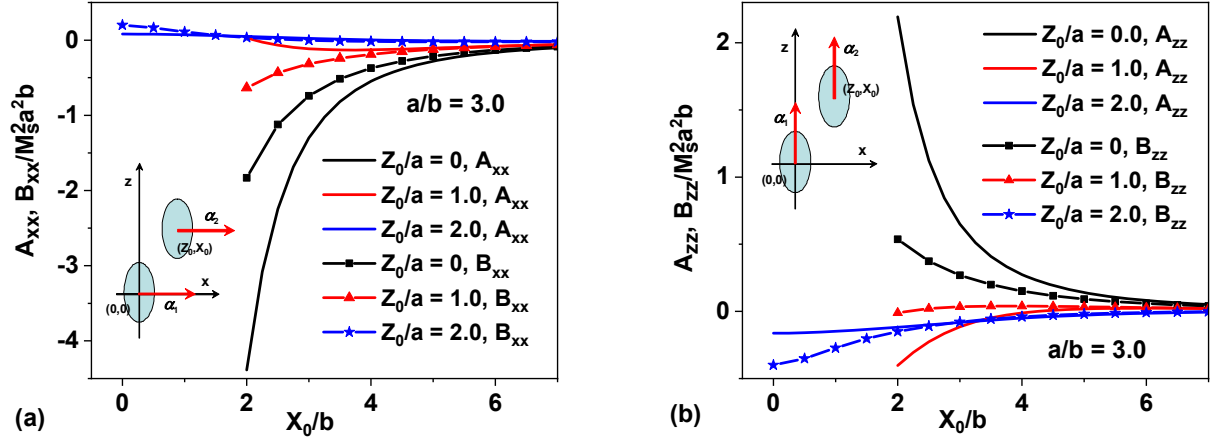

Fig. 3. Comparison of MD interaction matrix elements for spheroidal nanoparticles with aspect ratio $a/b = 3.0$ and spherical nanoparticles of equal volume. Dots show numerical results, solid lines are drawn for spherical approximation. The inserts show the arrangement of nanoparticles on the *XZ* plane and the corresponding orientation of their unit magnetization vectors.

To control the numerical results, it is useful to obtain approximate analytical expressions for the elements of the interaction matrix $B_{ij}$, which are valid for a pair of spheroidal nanoparticles oriented along the *Z* axis. For arbitrary position of the second nanoparticle, $\vec{R} = (X_0, Y_0, Z_0)$, by expanding the integrands in equations similar to Eq. (5) in powers of the parameter $\left(-2\vec{R}(\vec{r}_1 - \vec{r}_2) + (\vec{r}_1 - \vec{r}_2)^2\right)/R^2$, one can obtain the series expansions

$$B_{xx}^{(1)} = \frac{M_s^2 V^2}{R^3}\left(1 - 3n_x^2 + 3\delta^2\left(\frac{4}{5} - n_y^2 - 7n_x^2 n_z^2\right) + ...\right); \quad B_{yy}^{(1)} = \frac{M_s^2 V^2}{R^3}\left(1 - 3n_y^2 + 3\delta^2\left(\frac{4}{5} - n_x^2 - 7n_y^2 n_z^2\right) + ...\right);$$

$$B_{zz}^{(1)} = \frac{M_s^2 V^2}{R^3}\left(1 - 3n_z^2 + 3\delta^2\left(-\frac{3}{5} + 6n_z^2 - 7n_z^4\right) + ...\right) \quad B_{xy}^{(1)} = B_{yx}^{(1)} = -\frac{3M_s^2 V^2}{R^3} n_x n_y \left(1 - \delta^2(1 - 7n_z^2) + ...\right); \quad (15)$$

$$B_{xz}^{(1)} = B_{zx}^{(1)} = -\frac{3M_s^2 V^2}{R^3} n_x n_z \left(1 - \delta^2 \left(3 - 7n_z^2\right) + ...\right); \quad B_{yz}^{(1)} = B_{zy}^{(1)} = -\frac{3M_s^2 V^2}{R^3} n_y n_z \left(1 - \delta^2 \left(3 - 7n_z^2\right) + ...\right),$$

where $\delta^2 = (a^2 - b^2)/R^2$ turns out to be a small expansion parameter. Note that for spherical particles, $a = b$, all corrections of this kind vanish, as it should be. Therefore, Eqs. (15) give first order correction to spherical approximation, Eq. (12), for MD interaction matrix of oriented elongated spheroidal nanoparticles.

For completeness, in Fig. 4 the numerical results, obtained by means of Eqs. (6) – (9), are compared with series expansions, Eq. (15). As Fig. 4 shows, for the particle aspect ratios $a/b \leq 2.0$ formulas (15) describe the dependence of the components $B_{xx}$ and $B_{zz}$ on the reduced coordinate $X_0/b$ sufficiently accurately for values of the reduced coordinate $Z_0/a \geq 2.0$. However, in the range $0 \leq Z_0/a \leq 2.0$, significant deviations in the behavior of these dependences are observed for particles with aspect ratios $a/b \geq 1.5$. Thus, Eqs. (15) can be approximately used only for relatively small values of the particle aspect ratios, $a/b < 1.5$. However, the advantage of Eq. (15) is that they give continuous analytical representation for the matrix elements of the MD interaction matrix of oriented spheroidal nanoparticles in this range of particle aspect ratios.

Finally, Fig. 5 shows the maximum values of the basic components $B_{xx}$ and $B_{zz}$ of the MD interaction matrix reduced to the value $M_s^2 V$ as a function of particle aspect ratio. These diagonal components give the maximal values of the particle interaction energy density for the closest possible positions of oriented particles in the $XZ$ plane and for various characteristic directions of the unit magnetization vectors shown in the insets in Fig. 5. For example, the interaction matrix element $B_{xx}(2,0)$ corresponds to the case of particles located at the closest possible distance along the $X$ axis, when the space between the particle centers is given by $2b$, the unit magnetization vectors of the particles being directed along the $X$ axis. Similarly, for the same geometrical arrangements of the particles the matrix element $B_{zz}(2,0)$ corresponds to the case, when the unit magnetization vectors point along the $Z$ axis, and so on. Evidently, the case of $a/b = 1$ in Fig. 5 corresponds to spherical nanoparticles.

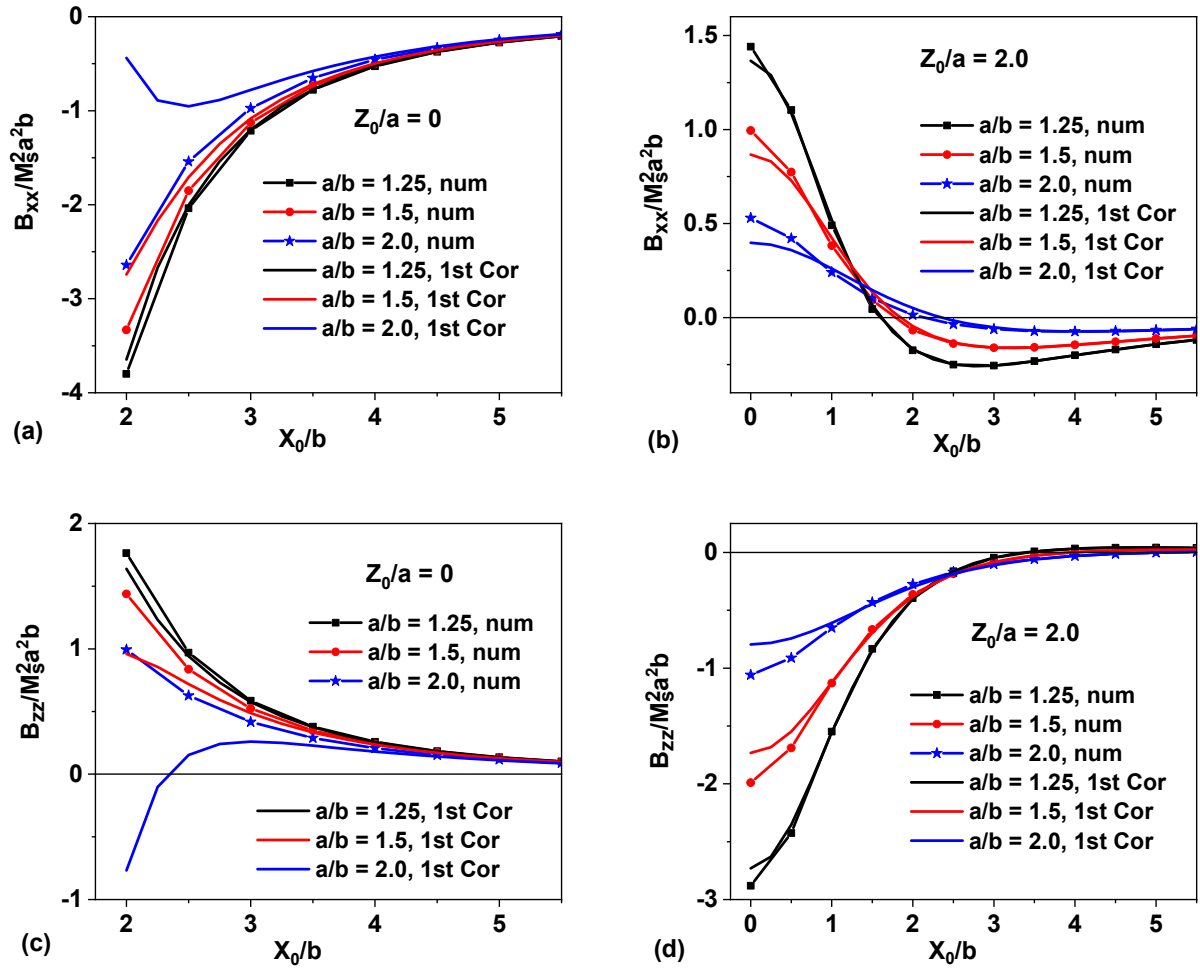

Fig. 4. Comparison of numerical results, obtained by means of Eqs. (6) and (8) for the reduced components $B_{xx}/M_s^2 a^2 b$ and $B_{zz}/M_s^2 a^2 b$ (dots), with first order corrections to spherical approximation, Eqs. (15) (solid curves), for MD interaction matrix of elongated spheroidal nanoparticles at various particle aspect ratios.

Finally, Fig. 5 shows the maximum values of the basic components $B_{xx}$ and $B_{zz}$ of the MD interaction matrix reduced to the value $M_s^2 V$ as a function of particle aspect ratio. These diagonal components give the maximal values of the particle interaction energy density for the closest possible positions of oriented particles in the $XZ$ plane and for various characteristic directions of the unit magnetization vectors shown in the insets in Fig. 5. For example, the interaction matrix element $B_{xx}(2,0)$ corresponds to the case of particles located at the closest possible distance along the $X$ axis, when the space between the particle centers is given by $2b$, the unit magnetization vectors of the particles being directed along the $X$ axis. Similarly, for the same geometrical arrangements of the particles the matrix element $B_{zz}(2,0)$ corresponds to the case, when the unit magnetization vectors point along the $Z$ axis, and so on. Evidently, the case of $a/b = 1$ in Fig. 5 corresponds to spherical nanoparticles.

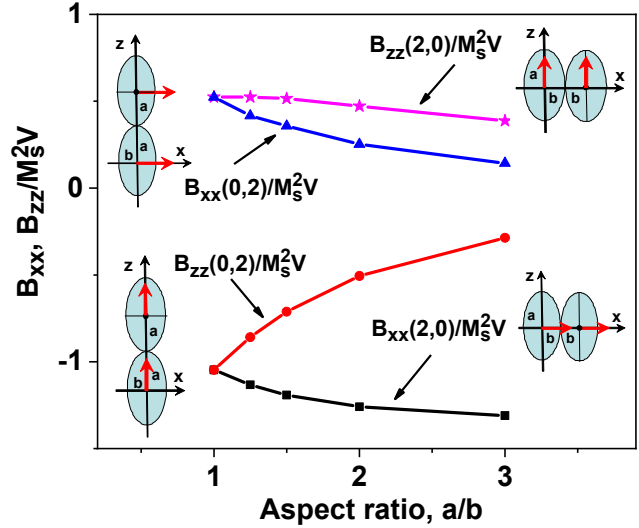

Fig. 5. Basic reduced diagonal elements $B_{xx}/M_s^2 V$ and $B_{zz}/M_s^2 V$ of MD interaction matrix of spheroidal nanoparticles as a function of particle aspect ratio for the closest particle arrangements in the $XZ$ plane. The insets show the arrangement of the particles and the corresponding orientation of their unit magnetization vectors, respectively.

It is interesting to note that the greatest dependence on the particle aspect ratio show the reduced matrix elements $B_{xx}(0,2)/M_s^2 V$ and $B_{zz}(0,2)/M_s^2 V$ which correspond to position of the second nanoparticle shifted along the Z axis to the closest distance 2a. Both of these matrix elements are decreased in the absolute value as a function of aspect ratio. The reduced matrix element $B_{xx}(2,0)/M_s^2 V$ is the only one whose absolute value increases as a function of aspect ratio.

To achieve sufficient accuracy in micromagnetic simulation, the numerical calculation of all nonzero matrix elements $B_{ij}$ as a function of the transverse coordinate was carried out up to the values $X_0/b$ = 10.0 with a step of 0.5. The calculation of these elements as a function of the longitudinal coordinate was made with the same step of 0.5 up to the values $Z_0/a$ = 6.0. In the micromagnetic calculations described below, for an arbitrary displacement of the second particle ($X_0/b$, $Z_0/a$) a three-point interpolation of the nearest tabular values for the corresponding matrix element was used. On the other hand, the matrix elements of the spherical approximation, Eq. (12), were used for sufficiently large distances between the particles, $X_0/b$ > 10.0 or $Z_0/a$ > 6.0.

**3. Quasi-static hysteresis loops**

Tables S1 – S4 for the MD interaction matrix elements $B_{xx}$, $B_{yy}$, $B_{zz}$ and $B_{xz}$ have recently been used [33] to calculate low-frequency hysteresis loops of dilute oriented clusters of spheroidal nanoparticles with random positions of particle centers within the cluster. However, in the calculations [33] the cluster filling density was relatively low, $\eta \leq 0.2$. To see better the effect of MD interaction energy on the properties of an assembly of elongated magnetic nanoparticles the quasi-static hysteresis loops are calculated for dilute oriented clusters of spheroidal magnetite nanoparticles with higher cluster filling density, up to $\eta \approx 0.4$. The cubic magnetic anisotropy constant and saturation magnetization of magnetite nanoparticles are assumed to be $K_c$ = - $10^5$ erg/cm$^3$ and $M_s$ = 450 emu/cm$^3$, respectively [34]. The calculations are carried out under the supposition that the directions of the particle cubic easy anisotropy axes are randomly oriented in space. At the same time, the long axes of spheroidal nanoparticles are supposed to be parallel to the Z axis.

Dynamics of the unit magnetization vector $\alpha_i$ of the *i*-th nanoparticle of the interacting assembly is described by the Landau-Lifshitz-Gilbert equation [31, 32]

$$\frac{\partial \vec{\alpha}_i}{\partial t} = -\gamma \left( \vec{\alpha}_i \times \vec{H}_{ef,i} \right) + \kappa \left( \vec{\alpha}_i \times \frac{\partial \vec{\alpha}_i}{\partial t} \right), \qquad i = 1, 2 \ldots, N_p, \qquad (16)$$

where $\kappa$ is the phenomenological damping constant, $\gamma$ is the gyromagnetic ratio, $N_p$ being the number of nanoparticles. The effective magnetic field acting on a separate nanoparticle is

$$\vec{H}_{ef,i} = -\frac{\partial W}{VM_s \partial \vec{\alpha}_i}. \qquad (17)$$

The total magnetic energy of the assembly $W = W_a + W_{sh} + W_Z + W_m$ is a sum of the cubic magneto-crystalline anisotropy energy $W_a$, shape anisotropy energy $W_{sh}$, Zeeman energy $W_Z$ of the particles in the external uniform magnetic field, and the energy of mutual magneto-dipole interaction of the particles $W_m$.

In a general situation, the cubic easy anisotropy axes of various magnetite nanoparticles oriented in space randomly. Therefore, the cubic magneto-crystalline anisotropy energy of the assembly is

$$W_a = K_c V \sum_{i=1}^{N_p} \left( (\vec{\alpha}_i \vec{e}_{1i})^2 (\vec{\alpha}_i \vec{e}_{2i})^2 + (\vec{\alpha}_i \vec{e}_{1i})^2 (\vec{\alpha}_i \vec{e}_{3i})^2 + (\vec{\alpha}_i \vec{e}_{2i})^2 (\vec{\alpha}_i \vec{e}_{3i})^2 \right). \qquad (18)$$

Here ($e_{1i}$, $e_{2i}$, $e_{3i}$) is a set of orthogonal unit vectors that determine the orientation of the easy anisotropy axes of the $i$-th nanoparticle of the assembly.

In this paper we consider an assembly of elongated nanoparticles of spheroidal shape with semiaxis ratio $a/b > 1$, oriented along the $Z$ axis. Therefore, in addition to the magneto-crystalline anisotropy energy, Eq. (18), there is also shape anisotropy energy contribution

$$W_{sh} = K_{sh} V \sum_{i=1}^{N_p} \left( \alpha_{i,x}^2 + \alpha_{i,y}^2 \right), \qquad (19)$$

where $K_{sh}$ is the shape anisotropy constant. The latter can be calculated as follows [35]

$$K_{sh} = M_s^2 (\pi - 3N_a/4); \qquad N_a = 2\pi \frac{1-\varepsilon^2}{\varepsilon^3} \left( \ln \frac{1+\varepsilon}{1-\varepsilon} - 2\varepsilon \right); \qquad \varepsilon = \sqrt{1 - (b/a)^2},$$

where $N_a$ is the demagnetizing factor along the long nanoparticle axis.

Zeeman energy $W_Z$ of the cluster in the external uniform magnetic field $H$ is given by

$$W_Z = -M_s V \sum_{i=1}^{N_p} (\vec{\alpha}_i \vec{H}). \tag{20}$$

The magneto-dipole interaction energy of a pair of single-domain spheroidal nanoparticles is described by Eq. (4). In the calculations performed the non-zero elements of interaction matrix $B_{xx}$, $B_{yy}$, $B_{zz}$ and $B_{xz}$ given in the Tables S1 – S4 were used for the reduced displacements of the second particle $X_0/b < 10.0$ and $Z_0/a < 6.0$. Otherwise, the formulas of spherical approximation, Eq. (12), were exploited.

Quasi-spherical clusters of a periodic structure were considered to study assemblies with increased particle filling density. In a periodic cluster each spheroidal particle was placed at the center of a rectangular cell with dimensions $L_x = L_y = 2\xi b$, $L_z = 2\xi a$, $\xi = 1.1 – 1.5$. The volume of a rectangular cell in which the nanoparticle is located is equal to $V_c = L_x^2 L_z$. Therefore, the filling density of a quasi-spherical cluster can be estimated as $\eta = V/V_c$, where $V = 4\pi ab^2/3$ is the volume of spheroidal nanoparticle.

Calculations of the quasi-static hysteresis loops were carried out for dilute assemblies of oriented clusters consisting $N_p = 60 – 90$ spheroidal nanoparticles. In a general case, the external magnetic field makes an angle $\theta$ with the orientation axis of the cluster. To obtain statistically reliable results, quasi-static hysteresis loops were averaged over an assembly of 20-30 independent cluster realizations. The hysteresis loop calculation began in a sufficiently strong external magnetic field sufficient for magnetic saturation of the cluster in a given direction. For given magnetic parameters of nanoparticles, the magnetic saturation field of the cluster ranged from 2000 to 3000 Oe, depending on the cluster filling density $\eta$. After saturation, the external magnetic field was successively decreased by a small value, $dH = 2 - 4$ Oe, and the dynamics of the unit magnetization vectors of nanoparticles was calculated in accordance with Eqs. (16), (17) until a stable equilibrium state is reached in a given external magnetic field. For clusters consisting of $N_p = 60 - 90$ nanoparticles it is usually required $10^4$ - $10^5$ iterations with a small time step to reach the equilibrium state. The latter, in accordance with Eq. (16), was assumed to be stable under the condition

$$\max_{(1 \leq i \leq N_p)} |\vec{\alpha}_i \times \vec{H}_{ef,i}| < 10^{-8}. \tag{21}$$

Fig. 6a shows quasi-static hysteresis loops for a dilute assembly of oriented clusters containing spheroidal nanoparticles with aspect ratios $a/b = 1.25$ and 3.0, respectively. The solid lines in Fig. 6a show hysteresis loops obtained using the exact MD interaction matrix, Eqs.

(6) – (9), whereas the dots show the corresponding loops obtained in the spherical approximation, Eqs. (12). The external uniform magnetic field is directed along the assembly orientation axis, $\theta = 0$. As Fig. 6a shows, for nanoparticles with aspect ratio $a/b = 1.25$ the hysteresis loop obtained in the spherical approximation practically coincides with that calculated using the exact MD interaction matrix. On the other hand, in agreement with the above discussion, a significant difference between exact and approximate hysteresis loops is found for assembly of spheroidal nanoparticles with aspect ratio $a/b = 3.0$, even for moderate cluster filling density $\eta = 0.155$. The loop difference increases as a function of the cluster filling density.

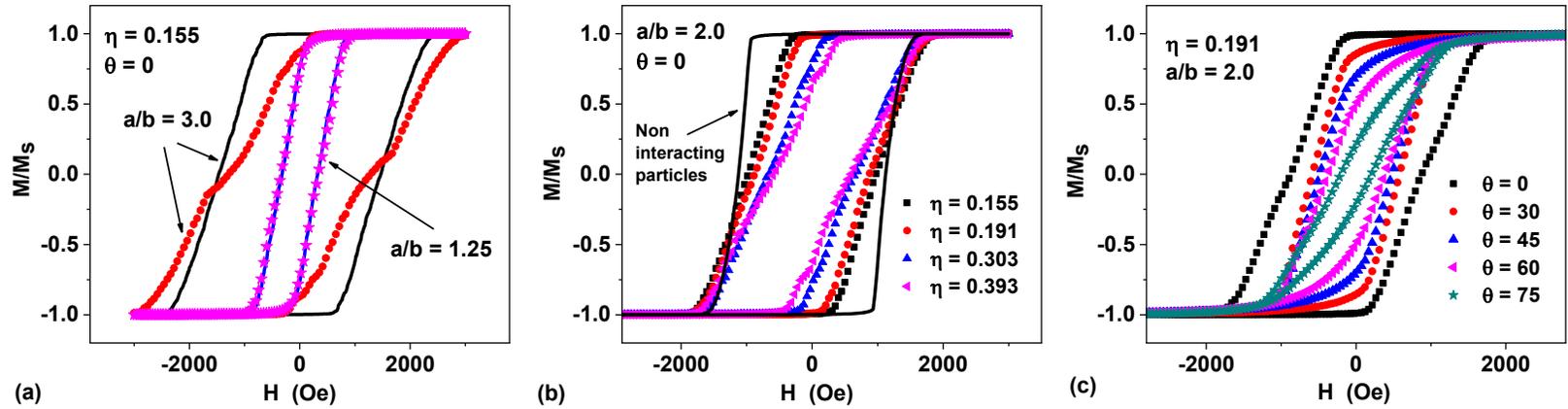

Fig. 6. a) Comparison of quasi-static hysteresis loops of a dilute assembly of oriented clusters of spheroidal nanoparticles with aspect ratios $a/b = 1.25$ and 3.0, respectively. Solid lines are calculated using the exact MD interaction matrix; dots correspond to the spherical approximation. b) Evolution of the quasi-static hysteresis loop as a function of cluster filling density at a fixed particle aspect ratio $a/b = 2.0$. c) Dependence of the quasi-static hysteresis loop on the angle $\theta$ the external magnetic field makes with the assembly orientation axis.

Fig. 6a also shows that the width of the quasi-static hysteresis loop significantly increases with an increase in the particle aspect ratio. Obviously, it is a consequence of an increase in the particle shape anisotropy energy as a function of the aspect ratio. Indeed, for a given particle saturation magnetization $M_s$ = 450 emu/cm$^3$, the shape anisotropy constant increases from $K_{sh}$ = 1.09×10$^5$ erg/cm$^3$ at particle aspect ratio $a/b$ = 1.25 to $K_{sh}$ = 4.29×10$^5$ erg/cm$^3$ at aspect ratio $a/b$ = 3.0.

More interesting is the dependence of the hysteresis loop shape of dilute assemblies of oriented clusters on the cluster filling density $\eta$ for a fixed particle aspect ratio. As Fig. 6b shows, with an increase in filling density the width of the quasi-static hysteresis loop appreciably decreases, whereas its slope increases. This leads to a decrease in the remanent magnetization and the coercive force of the assembly. Obviously, these effects are associated with an increase in the cluster MD interaction energy as a function of the filling density. These changes are found to be more noticeable for particles with a small aspect ratio, when the orienting role of the shape anisotropy energy is less pronounced. The angular dependence of the quasi-static hysteresis loops of a dilute assembly of oriented clusters is shown in Fig. 6c.

It is worth noting that at a fixed filling density $\eta$ no significant dependence of the quasi-static hysteresis loops on the number $N_p$ = 60 – 90 of particles in the cluster was found.

## 4. Conclusions

Elongated magnetic nanoparticles are of interest for various areas of nanotechnology [27-30]. In particularly, it was shown recently [33,36,37] that orientation of assembly of elongated magnetic nanoparticles leads to increased value of specific absorption rate of the assembly in alternating magnetic field. Experimentally, the orientation of an assembly of elongated magnetic nanoparticles can be performed in a viscous liquid [36, 37], in a sufficiently strong static external magnetic field. It was also reported [28,38] that the orientation of elongated nanoparticles can occur in viscous liquid or biological media under the influence of applied alternating magnetic field. In all these cases, to correctly estimate the influence of the MD interaction energy on the properties of a dense oriented assembly of elongated magnetic nanoparticles, it is necessary to use the elements of the exact MD interaction matrix calculated in this work.

**CRediT authorship contribution statement N.A. Usov:** Conceptualization, Software, Validation, Writing – review & editing.

**Declaration of Competing Interest** The author declare that he has no known competing financial interests or personal relationships that could have appeared to influence the work reported in this paper.

**Acknowledgment**

The author gratefully acknowledges the financial support of the Ministry of Science and Higher Education of the Russian Federation in the framework of Increase Competitiveness Program of NUST «MISIS», contract № K2-2019-012.

# Supplementary Information for the paper

Magnetostatic interaction in oriented assembly of elongated nanoparticles

By N. A. Usov[1,2]

[1] National University of Science and Technology «MISiS», 119049, Moscow, Russia
[2] Pushkov Institute of Terrestrial Magnetism, Ionosphere and Radio Wave Propagation, Russian Academy of Sciences, IZMIRAN, 108480, Troitsk, Moscow, Russia

**Table S1, a/b = 1.25**

$B_{xx}/M_s^2 a^2 b$

| $Z_0/a$ | 0 | 0.5 | 1.0 | 1.5 | 2.0 | 2.5 | 3.0 | 3.5 | 4.0 | 4.5 | 5.0 | 5.5 | 6.0 |
|---|---|---|---|---|---|---|---|---|---|---|---|---|---|
| $X_0/b$ | | | | | | | | | | | | | |
| 0 | | | | | 1.44096 | 0.66852 | 0.36839 | 0.22554 | 0.14843 | 0.103 | 0.07445 | 0.05559 | 0.04262 |
| 0.5 | | | | | 1.10391 | 0.57784 | 0.33544 | 0.21116 | 0.14132 | 0.09916 | 0.07222 | 0.05422 | 0.04174 |
| 1 | | | | | 0.49039 | 0.37144 | 0.25263 | 0.17306 | 0.12188 | 0.08842 | 0.0659 | 0.05031 | 0.03921 |
| 1.5 | | | | | 0.04506 | 0.16304 | 0.15379 | 0.12303 | 0.0948 | 0.07285 | 0.05648 | 0.04436 | 0.03531 |
| 2 | -3.79934 | -3.10953 | -1.73674 | -0.68494 | -0.17434 | 0.01322 | 0.06656 | 0.07306 | 0.06547 | 0.05504 | 0.04528 | 0.03708 | 0.03043 |
| 2.5 | -2.03839 | -1.76517 | -1.16305 | -0.60312 | -0.25001 | -0.0732 | 0.00281 | 0.03087 | 0.03824 | 0.03737 | 0.03362 | 0.02921 | 0.02502 |
| 3 | -1.21257 | -1.08918 | -0.79737 | -0.48777 | -0.2553 | -0.11341 | -0.03751 | -6.16E-4 | 0.01559 | 0.0215 | 0.02255 | 0.02143 | 0.01948 |
| 3.5 | -0.77702 | -0.71543 | -0.56224 | -0.38381 | -0.23217 | -0.12566 | -0.05955 | -0.02182 | -0.00167 | 0.00834 | 0.01277 | 0.01422 | 0.01416 |
| 4 | -0.52666 | -0.49338 | -0.40752 | -0.30046 | -0.20078 | -0.12305 | -0.06915 | -0.03467 | -0.01385 | -0.00189 | 0.00463 | 0.00791 | 0.00931 |
| 4.5 | -0.37293 | -0.35376 | -0.30296 | -0.23627 | -0.16975 | -0.11353 | -0.07106 | -0.04138 | -0.02177 | -0.00936 | -0.0018 | 0.00263 | 0.00506 |
| 5 | -0.27348 | -0.26185 | -0.23038 | -0.18742 | -0.14221 | -0.10151 | -0.0686 | -0.04392 | -0.02641 | -0.01449 | -0.00664 | -0.00161 | 0.00149 |
| 5.5 | -0.20639 | -0.19902 | -0.17875 | -0.15022 | -0.1189 | -0.08926 | -0.06392 | -0.0438 | -0.02867 | -0.01774 | -0.01009 | -0.00488 | -0.00141 |
| 6 | -0.15951 | -0.15467 | -0.14117 | -0.1217 | -0.0996 | -0.07781 | -0.05833 | -0.0421 | -0.02928 | -0.01956 | -0.01242 | -0.00729 | -0.0037 |
| 6.5 | -0.1258 | -0.12251 | -0.11325 | -0.09965 | -0.08377 | -0.06758 | -0.05256 | -0.03953 | -0.02881 | -0.02034 | -0.01384 | -0.00899 | -0.00543 |
| 7 | -0.10094 | -0.09864 | -0.09214 | -0.08242 | -0.07082 | -0.05866 | -0.04701 | -0.03657 | -0.02767 | -0.02038 | -0.01459 | -0.0101 | -0.00669 |
| 7.5 | -0.08221 | -0.08057 | -0.07589 | -0.06881 | -0.0602 | -0.05097 | -0.04189 | -0.0335 | -0.02614 | -0.01992 | -0.01483 | -0.01075 | -0.00756 |
| 8 | -0.06783 | -0.06664 | -0.06321 | -0.05796 | -0.05148 | -0.04439 | -0.03726 | -0.03051 | -0.02443 | -0.01915 | -0.0147 | -0.01105 | -0.00812 |
| 8.5 | -0.05662 | -0.05573 | -0.05317 | -0.04922 | -0.04427 | -0.03877 | -0.03313 | -0.02768 | -0.02265 | -0.01818 | -0.01433 | -0.01109 | -0.00842 |
| 9 | -0.04774 | -0.04707 | -0.04513 | -0.04212 | -0.03829 | -0.03398 | -0.02949 | -0.02506 | -0.0209 | -0.01712 | -0.01379 | -0.01094 | -0.00853 |
| 9.5 | -0.04063 | -0.04012 | -0.03863 | -0.03629 | -0.0333 | -0.02989 | -0.02628 | -0.02267 | -0.01921 | -0.01602 | -0.01316 | -0.01065 | -0.00849 |
| 10 | -0.03486 | -0.03446 | -0.0333 | -0.03147 | -0.02911 | -0.02639 | -0.02347 | -0.02051 | -0.01763 | -0.01493 | -0.01246 | -0.01026 | -0.00834 |

**$B_{yy}/M_s^2 a^2 b$**

| $Z_0/a$ | 0 | 0.5 | 1.0 | 1.5 | 2.0 | 2.5 | 3.0 | 3.5 | 4.0 | 4.5 | 5.0 | 5.5 | 6.0 |
|---|---|---|---|---|---|---|---|---|---|---|---|---|---|
| $X_0/b$ | | | | | | | | | | | | | |
| 0 | | | | | 1.44096 | 0.66852 | 0.36839 | 0.22554 | 0.14843 | 0.103 | 0.07445 | 0.05559 | 0.04262 |
| 0.5 | | | | | 1.32202 | 0.63739 | 0.35721 | 0.22069 | 0.14604 | 0.10171 | 0.07371 | 0.05513 | 0.04232 |
| 1 | | | | | 1.05846 | 0.55829 | 0.32703 | 0.20717 | 0.13925 | 0.098 | 0.07154 | 0.0538 | 0.04147 |
| 1.5 | | | | | 0.78801 | 0.46013 | 0.28574 | 0.18761 | 0.12908 | 0.09231 | 0.06816 | 0.05169 | 0.0401 |
| 2 | 2.03567 | 1.82658 | 1.35834 | 0.89885 | 0.57089 | 0.36571 | 0.24142 | 0.16512 | 0.11684 | 0.08524 | 0.06386 | 0.04897 | 0.03831 |
| 2.5 | 1.06849 | 0.99018 | 0.80092 | 0.58896 | 0.4126 | 0.28563 | 0.19966 | 0.14232 | 0.10378 | 0.07741 | 0.05898 | 0.04581 | 0.0362 |
| 3 | 0.6273 | 0.59317 | 0.50633 | 0.39988 | 0.3011 | 0.22197 | 0.1632 | 0.12097 | 0.0909 | 0.06938 | 0.05381 | 0.04239 | 0.03387 |
| 3.5 | 0.39861 | 0.38197 | 0.33808 | 0.28072 | 0.22309 | 0.17294 | 0.13277 | 0.10197 | 0.07884 | 0.06157 | 0.04863 | 0.03887 | 0.03143 |
| 4 | 0.26864 | 0.25979 | 0.23586 | 0.20308 | 0.16811 | 0.13568 | 0.10803 | 0.08562 | 0.06797 | 0.05425 | 0.04362 | 0.03538 | 0.02896 |
| 4.5 | 0.18946 | 0.18443 | 0.17055 | 0.15086 | 0.12887 | 0.10742 | 0.08818 | 0.07184 | 0.0584 | 0.04757 | 0.03892 | 0.03203 | 0.02654 |
| 5 | 0.13853 | 0.13551 | 0.12704 | 0.1147 | 0.10042 | 0.08592 | 0.07235 | 0.06036 | 0.05013 | 0.0416 | 0.03459 | 0.02887 | 0.02421 |
| 5.5 | 0.10432 | 0.10241 | 0.09702 | 0.089 | 0.07946 | 0.06944 | 0.05974 | 0.05087 | 0.04305 | 0.03634 | 0.03068 | 0.02594 | 0.022 |
| 6 | 0.08049 | 0.07924 | 0.07568 | 0.0703 | 0.06376 | 0.0567 | 0.04967 | 0.04304 | 0.03704 | 0.03175 | 0.02718 | 0.02327 | 0.01995 |
| 6.5 | 0.06339 | 0.06255 | 0.06013 | 0.05642 | 0.05182 | 0.04675 | 0.04158 | 0.03659 | 0.03195 | 0.02777 | 0.02407 | 0.02085 | 0.01806 |
| 7 | 0.05081 | 0.05022 | 0.04853 | 0.04591 | 0.04261 | 0.03891 | 0.03506 | 0.03126 | 0.02765 | 0.02433 | 0.02133 | 0.01867 | 0.01634 |
| 7.5 | 0.04135 | 0.04093 | 0.03972 | 0.03783 | 0.03542 | 0.03267 | 0.02976 | 0.02684 | 0.02401 | 0.02136 | 0.01893 | 0.01673 | 0.01477 |
| 8 | 0.03409 | 0.03379 | 0.03291 | 0.03151 | 0.02972 | 0.02765 | 0.02543 | 0.02316 | 0.02093 | 0.0188 | 0.01682 | 0.015 | 0.01336 |
| 8.5 | 0.02844 | 0.02822 | 0.02756 | 0.02652 | 0.02516 | 0.02358 | 0.02187 | 0.02009 | 0.01831 | 0.0166 | 0.01498 | 0.01347 | 0.01208 |
| 9 | 0.02397 | 0.0238 | 0.0233 | 0.02251 | 0.02148 | 0.02026 | 0.01891 | 0.01751 | 0.01609 | 0.0147 | 0.01336 | 0.01211 | 0.01094 |
| 9.5 | 0.02039 | 0.02026 | 0.01988 | 0.01927 | 0.01847 | 0.01751 | 0.01645 | 0.01533 | 0.01419 | 0.01305 | 0.01195 | 0.0109 | 0.00992 |
| 10 | 0.01749 | 0.01739 | 0.01709 | 0.01662 | 0.01599 | 0.01523 | 0.01439 | 0.01349 | 0.01256 | 0.01162 | 0.01071 | 0.00983 | 0.009 |

**$B_{zz}/M_s^2 a^2 b$**

| $Z_0/a$ | 0 | 0.5 | 1.0 | 1.5 | 2.0 | 2.5 | 3.0 | 3.5 | 4.0 | 4.5 | 5.0 | 5.5 | 6.0 |
|---|---|---|---|---|---|---|---|---|---|---|---|---|---|
| $X_0/b$ | | | | | | | | | | | | | |
| 0 | | | | | -2.88161 | -1.33697 | -0.73676 | -0.45107 | -0.29684 | -0.206 | -0.1489 | -0.11118 | -0.08524 |
| 0.5 | | | | | -2.42577 | -1.21518 | -0.69263 | -0.43184 | -0.28735 | -0.20086 | -0.14593 | -0.10936 | -0.08406 |
| 1 | | | | | -1.54881 | -0.92971 | -0.57964 | -0.38022 | -0.26113 | -0.18641 | -0.13744 | -0.10411 | -0.08068 |
| 1.5 | | | | | -0.83307 | -0.62316 | -0.43952 | -0.31064 | -0.22388 | -0.16515 | -0.12464 | -0.09605 | -0.07541 |
| 2 | 1.76391 | 1.28313 | 0.37843 | -0.21392 | -0.39656 | -0.37893 | -0.30799 | -0.23818 | -0.18231 | -0.14028 | -0.10915 | -0.08604 | -0.06874 |
| 2.5 | 0.97 | 0.77506 | 0.36215 | 0.01416 | -0.1626 | -0.21244 | -0.20248 | -0.1732 | -0.14202 | -0.11478 | -0.0926 | -0.07502 | -0.06121 |
| 3 | 0.58531 | 0.49604 | 0.29105 | 0.0879 | -0.0458 | -0.10856 | -0.12569 | -0.12036 | -0.10649 | -0.09089 | -0.07636 | -0.06381 | -0.05335 |
| 3.5 | 0.37844 | 0.33348 | 0.22417 | 0.10309 | 0.00908 | -0.04728 | -0.07322 | -0.08015 | -0.07717 | -0.06991 | -0.0614 | -0.05309 | -0.04559 |
| 4 | 0.25803 | 0.23359 | 0.17167 | 0.09738 | 0.03267 | -0.01263 | -0.03888 | -0.05096 | -0.05412 | -0.05236 | -0.04826 | -0.0433 | -0.03827 |
| 4.5 | 0.18347 | 0.16934 | 0.13242 | 0.08541 | 0.04087 | 0.00611 | -0.01713 | -0.03046 | -0.03663 | -0.03821 | -0.03712 | -0.03466 | -0.0316 |
| 5 | 0.13495 | 0.12635 | 0.10335 | 0.07272 | 0.04179 | 0.0156 | -0.00376 | -0.01644 | -0.02372 | -0.02711 | -0.02796 | -0.02725 | -0.0257 |
| 5.5 | 0.10207 | 0.09661 | 0.08173 | 0.06122 | 0.03944 | 0.01982 | 0.00418 | -0.00707 | -0.01438 | -0.0186 | -0.02058 | -0.02106 | -0.02059 |
| 6 | 0.07903 | 0.07542 | 0.06548 | 0.0514 | 0.03585 | 0.02111 | 0.00866 | -9.47E-4 | -0.00776 | -0.01219 | -0.01476 | -0.01597 | -0.01626 |
| 6.5 | 0.06241 | 0.05996 | 0.05313 | 0.04323 | 0.03195 | 0.02083 | 0.01098 | 0.00294 | -0.00314 | -0.00743 | -0.01023 | -0.01186 | -0.01263 |
| 7 | 0.05012 | 0.04842 | 0.0436 | 0.03651 | 0.0282 | 0.01975 | 0.01196 | 0.00531 | 1.79E-5 | -0.00395 | -0.00675 | -0.00857 | -0.00964 |
| 7.5 | 0.04086 | 0.03964 | 0.03617 | 0.03098 | 0.02479 | 0.0183 | 0.01213 | 0.00667 | 0.00213 | -0.00144 | -0.0041 | -0.00598 | -0.00721 |

| 8 | 0.03374 | 0.03285 | 0.0303 | 0.02644 | 0.02176 | 0.01674 | 0.01183 | 0.00735 | 0.00349 | 3.43E-4 | -0.00212 | -0.00395 | -0.00524 |
| 8.5 | 0.02818 | 0.02752 | 0.02561 | 0.02271 | 0.01911 | 0.01519 | 0.01126 | 0.00759 | 0.00434 | 0.00159 | -6.46E-4 | -0.00237 | -0.00367 |
| 9 | 0.02378 | 0.02328 | 0.02183 | 0.0196 | 0.01682 | 0.01373 | 0.01057 | 0.00755 | 0.00481 | 0.00243 | 4.33E-4 | -0.00117 | -0.00241 |
| 9.5 | 0.02024 | 0.01986 | 0.01875 | 0.01702 | 0.01484 | 0.01238 | 0.00983 | 0.00734 | 0.00503 | 0.00297 | 0.00121 | -2.54E-4 | -0.00143 |
| 10 | 0.01737 | 0.01708 | 0.01621 | 0.01485 | 0.01313 | 0.01116 | 0.00908 | 0.00702 | 0.00508 | 0.0033 | 0.00175 | 4.32E-4 | -6.52E-4 |

$B_{xz}/M_s^2 a^2 b$

| $Z_0/a$ | 0 | 0.5 | 1.0 | 1.5 | 2.0 | 2.5 | 3.0 | 3.5 | 4.0 | 4.5 | 5.0 | 5.5 | 6.0 |
|---|---|---|---|---|---|---|---|---|---|---|---|---|---|
| $X_0/b$ | | | | | | | | | | | | | |
| 0 | | | | | 0 | 0 | 0 | 0 | 0 | 0 | 0 | 0 | 0 |
| 0.5 | | | | | -0.88194 | -0.32752 | -0.14978 | -0.07832 | -0.04498 | -0.02769 | -0.01799 | -0.0122 | -0.00856 |
| 1 | | | | | -1.18901 | -0.5201 | -0.25737 | -0.14063 | -0.08298 | -0.05201 | -0.03421 | -0.02341 | -0.01654 |
| 1.5 | | | | | -1.07367 | -0.55978 | -0.3066 | -0.17823 | -0.10948 | -0.07052 | -0.0473 | -0.03283 | -0.02346 |
| 2 | 0 | -1.23656 | -1.6004 | -1.27813 | -0.83158 | -0.50573 | -0.30724 | -0.19147 | -0.12337 | -0.08222 | -0.05653 | -0.03997 | -0.02898 |
| 2.5 | 0 | -0.59162 | -0.85776 | -0.79851 | -0.60506 | -0.41743 | -0.27893 | -0.18634 | -0.12632 | -0.0874 | -0.06182 | -0.04466 | -0.03294 |
| 3 | 0 | -0.31382 | -0.49112 | -0.50828 | -0.43083 | -0.32884 | -0.23874 | -0.17023 | -0.12134 | -0.08727 | -0.06361 | -0.04706 | -0.03536 |
| 3.5 | 0 | -0.18025 | -0.2975 | -0.33213 | -0.3062 | -0.25332 | -0.19736 | -0.14923 | -0.11155 | -0.08334 | -0.06263 | -0.04749 | -0.0364 |
| 4 | 0 | -0.11024 | -0.18905 | -0.22315 | -0.21934 | -0.19353 | -0.15998 | -0.12739 | -0.09948 | -0.07707 | -0.05967 | -0.04638 | -0.03628 |
| 4.5 | 0 | -0.07094 | -0.12513 | -0.15402 | -0.1591 | -0.14783 | -0.12842 | -0.10696 | -0.08688 | -0.06962 | -0.05547 | -0.04417 | -0.03526 |
| 5 | 0 | -0.04758 | -0.08575 | -0.10901 | -0.11709 | -0.11345 | -0.10273 | -0.08895 | -0.07483 | -0.06186 | -0.05063 | -0.04126 | -0.0336 |
| 5.5 | 0 | -0.03305 | -0.06054 | -0.07893 | -0.08746 | -0.08769 | -0.08222 | -0.07362 | -0.06389 | -0.05431 | -0.04558 | -0.03797 | -0.03152 |
| 6 | 0 | -0.02364 | -0.04387 | -0.05834 | -0.0663 | -0.06838 | -0.06601 | -0.06083 | -0.05426 | -0.04731 | -0.04062 | -0.03454 | -0.0292 |
| 6.5 | 0 | -0.01734 | -0.03251 | -0.04394 | -0.05097 | -0.05381 | -0.05324 | -0.0503 | -0.04597 | -0.041 | -0.03595 | -0.03116 | -0.0268 |
| 7 | 0 | -0.013 | -0.02458 | -0.03366 | -0.0397 | -0.04274 | -0.04319 | -0.0417 | -0.03892 | -0.03542 | -0.03166 | -0.02793 | -0.02442 |
| 7.5 | 0 | -0.00993 | -0.0189 | -0.02617 | -0.03131 | -0.03426 | -0.03526 | -0.03468 | -0.03298 | -0.03057 | -0.0278 | -0.02492 | -0.02211 |
| 8 | 0 | -0.00771 | -0.01477 | -0.02062 | -0.02498 | -0.02772 | -0.02897 | -0.02896 | -0.02799 | -0.02638 | -0.02436 | -0.02217 | -0.01994 |
| 8.5 | 0 | -0.00608 | -0.0117 | -0.01646 | -0.02014 | -0.02262 | -0.02395 | -0.02429 | -0.02382 | -0.02277 | -0.02134 | -0.01968 | -0.01794 |
| 9 | 0 | -0.00486 | -0.00938 | -0.01329 | -0.01639 | -0.0186 | -0.01993 | -0.02046 | -0.02033 | -0.01969 | -0.01868 | -0.01745 | -0.01609 |
| 9.5 | 0 | -0.00392 | -0.00761 | -0.01083 | -0.01346 | -0.01542 | -0.01669 | -0.01732 | -0.01741 | -0.01706 | -0.01638 | -0.01546 | -0.01442 |
| 10 | 0 | -0.00321 | -0.00623 | -0.00892 | -0.01116 | -0.01287 | -0.01406 | -0.01474 | -0.01496 | -0.01481 | -0.01437 | -0.01371 | -0.01291 |

**Table S2, a/b = 1.5**

$B_{xx}/M_s^2 a^2 b$

| $Z_0/a$ | 0 | 0.5 | 1.0 | 1.5 | 2.0 | 2.5 | 3.0 | 3.5 | 4.0 | 4.5 | 5.0 | 5.5 | 6.0 |
|---|---|---|---|---|---|---|---|---|---|---|---|---|---|
| $X_0/b$ | | | | | | | | | | | | | |
| 0 | | | | | 0.99500 | 0.42538 | 0.22656 | 0.13622 | 0.08868 | 0.06101 | 0.04395 | 0.0327 | 0.02500 |
| 0.5 | | | | | 0.77352 | 0.37788 | 0.21089 | 0.12972 | 0.08556 | 0.05945 | 0.04300 | 0.03212 | 0.02463 |

| $X_0/b$ | 0 | 0.5 | 1.0 | 1.5 | 2.0 | 2.5 | 3.0 | 3.5 | 4.0 | 4.5 | 5.0 | 5.5 | 6.0 |
|---|---|---|---|---|---|---|---|---|---|---|---|---|---|
| 1 | | | | | 0.38304 | 0.26617 | 0.17012 | 0.11200 | 0.07683 | 0.05474 | 0.04028 | 0.03046 | 0.02357 |
| 1.5 | | | | | 0.09249 | 0.14558 | 0.11825 | 0.08744 | 0.06413 | 0.04766 | 0.03610 | 0.02786 | 0.02189 |
| 2 | -3.33031 | -2.69019 | -1.41664 | -0.47380 | -0.06761 | 0.04959 | 0.06819 | 0.06099 | 0.04949 | 0.03915 | 0.03090 | 0.02456 | 0.01972 |
| 2.5 | -1.85089 | -1.57012 | -0.96528 | -0.43601 | -0.13893 | -0.01431 | 0.02724 | 0.03649 | 0.03480 | 0.03012 | 0.02519 | 0.02083 | 0.01721 |
| 3 | -1.12738 | -0.99091 | -0.67829 | -0.36883 | -0.16079 | -0.05133 | -0.00258 | 0.01607 | 0.02142 | 0.02138 | 0.01940 | 0.01693 | 0.01452 |
| 3.5 | -0.73432 | -0.66244 | -0.48972 | -0.30206 | -0.15822 | -0.06956 | -0.02227 | 3.92E-04 | 0.01010 | 0.01347 | 0.01390 | 0.01308 | 0.01179 |
| 4 | -0.50351 | -0.46310 | -0.36231 | -0.24466 | -0.14505 | -0.07597 | -0.03403 | -0.01082 | 0.00110 | 0.00669 | 0.00893 | 0.00945 | 0.00915 |
| 4.5 | -0.35957 | -0.33559 | -0.27399 | -0.19796 | -0.12841 | -0.07555 | -0.04009 | -0.01827 | -0.00569 | 0.00116 | 0.00463 | 0.00618 | 0.00667 |
| 5 | -0.26536 | -0.25047 | -0.21133 | -0.16080 | -0.11166 | -0.07148 | -0.04231 | -0.02281 | -0.01054 | -0.00317 | 0.00104 | 0.00332 | 0.00443 |
| 5.5 | -0.20123 | -0.19163 | -0.16588 | -0.13145 | -0.09627 | -0.06575 | -0.04210 | -0.02520 | -0.01379 | -0.00642 | -0.00185 | 8.92E-04 | 0.00245 |
| 6 | -0.15612 | -0.14971 | -0.13228 | -0.10827 | -0.08272 | -0.05947 | -0.04045 | -0.02607 | -0.01579 | -0.00875 | -0.00409 | -0.00111 | 7.50E-04 |
| 6.5 | -0.12349 | -0.11909 | -0.10698 | -0.08989 | -0.07108 | -0.05326 | -0.03801 | -0.02592 | -0.01685 | -0.01033 | -0.00578 | -0.00271 | -6.75E-04 |
| 7 | -0.09933 | -0.09623 | -0.08761 | -0.07522 | -0.06119 | -0.04743 | -0.03521 | -0.02512 | -0.01722 | -0.01130 | -0.00700 | -0.00395 | -0.00184 |
| 7.5 | -0.08106 | -0.07883 | -0.07257 | -0.06343 | -0.05283 | -0.04213 | -0.03231 | -0.02392 | -0.01711 | -0.01181 | -0.00782 | -0.00489 | -0.00277 |
| 8 | -0.06699 | -0.06536 | -0.06074 | -0.05388 | -0.04577 | -0.03739 | -0.02948 | -0.02250 | -0.01667 | -0.01198 | -0.00833 | -0.00556 | -0.00350 |
| 8.5 | -0.05600 | -0.05478 | -0.05130 | -0.04609 | -0.03982 | -0.03320 | -0.02679 | -0.02100 | -0.01601 | -0.01190 | -0.00860 | -0.00602 | -0.00405 |
| 9 | -0.04727 | -0.04635 | -0.04370 | -0.03968 | -0.03478 | -0.02951 | -0.02430 | -0.01948 | -0.01523 | -0.01163 | -0.00867 | -0.00630 | -0.00444 |
| 9.5 | -0.04027 | -0.03956 | -0.03751 | -0.03438 | -0.03051 | -0.02628 | -0.02203 | -0.01800 | -0.01438 | -0.01124 | -0.00861 | -0.00645 | -0.00471 |
| 10 | -0.03458 | -0.03402 | -0.03242 | -0.02996 | -0.02687 | -0.02346 | -0.01996 | -0.01660 | -0.01351 | -0.01078 | -0.00844 | -0.00648 | -0.00488 |

$B_{yy}/M_s^2 a^2 b$

| $Z_0/a$ \ $X_0/b$ | 0 | 0.5 | 1.0 | 1.5 | 2.0 | 2.5 | 3.0 | 3.5 | 4.0 | 4.5 | 5.0 | 5.5 | 6.0 |
|---|---|---|---|---|---|---|---|---|---|---|---|---|---|
| 0 | | | | | 0.995 | 0.42538 | 0.22656 | 0.13622 | 0.08868 | 0.0611 | 0.04395 | 0.0327 | 0.025 |
| 0.5 | | | | | 0.91624 | 0.40913 | 0.22126 | 0.13404 | 0.08763 | 0.06054 | 0.04363 | 0.0325 | 0.02488 |
| 1 | | | | | 0.7463 | 0.36704 | 0.20666 | 0.12784 | 0.08462 | 0.05893 | 0.0427 | 0.03194 | 0.02452 |
| 1.5 | | | | | 0.57286 | 0.31294 | 0.18594 | 0.11858 | 0.07999 | 0.05641 | 0.04123 | 0.03103 | 0.02394 |
| 2 | 1.89215 | 1.6731 | 1.18868 | 0.73217 | 0.43042 | 0.25844 | 0.16262 | 0.10748 | 0.07421 | 0.05318 | 0.03932 | 0.02984 | 0.02316 |
| 2.5 | 1.01434 | 0.9256 | 0.71627 | 0.4938 | 0.32257 | 0.20977 | 0.13946 | 0.09568 | 0.06779 | 0.04947 | 0.03707 | 0.02842 | 0.02223 |
| 3 | 0.60359 | 0.56274 | 0.46164 | 0.34431 | 0.24335 | 0.16902 | 0.11812 | 0.08405 | 0.06115 | 0.04551 | 0.0346 | 0.02683 | 0.02117 |
| 3.5 | 0.38701 | 0.3663 | 0.31317 | 0.24728 | 0.18561 | 0.13606 | 0.09936 | 0.07316 | 0.05464 | 0.04148 | 0.03203 | 0.02514 | 0.02002 |
| 4 | 0.26246 | 0.25114 | 0.22127 | 0.18231 | 0.14339 | 0.10986 | 0.08335 | 0.06332 | 0.04848 | 0.03755 | 0.02945 | 0.0234 | 0.01883 |
| 4.5 | 0.18594 | 0.17936 | 0.16162 | 0.13756 | 0.11225 | 0.08919 | 0.06993 | 0.05464 | 0.04282 | 0.0338 | 0.02692 | 0.02166 | 0.01761 |
| 5 | 0.13642 | 0.13239 | 0.12137 | 0.10594 | 0.08903 | 0.0729 | 0.05879 | 0.0471 | 0.03771 | 0.03031 | 0.02451 | 0.01997 | 0.0164 |
| 5.5 | 0.10298 | 0.10042 | 0.09329 | 0.08308 | 0.0715 | 0.06003 | 0.0496 | 0.04062 | 0.03317 | 0.02711 | 0.02224 | 0.01834 | 0.01522 |
| 6 | 0.07962 | 0.07792 | 0.07316 | 0.0662 | 0.0581 | 0.04981 | 0.04202 | 0.03509 | 0.02916 | 0.02421 | 0.02013 | 0.0168 | 0.01409 |
| 6.5 | 0.0628 | 0.06165 | 0.05838 | 0.05352 | 0.04772 | 0.04164 | 0.03576 | 0.03039 | 0.02566 | 0.02161 | 0.0182 | 0.01536 | 0.013 |
| 7 | 0.0504 | 0.04959 | 0.04729 | 0.04382 | 0.0396 | 0.03508 | 0.03059 | 0.02639 | 0.02261 | 0.01929 | 0.01644 | 0.01402 | 0.01198 |
| 7.5 | 0.04105 | 0.04048 | 0.03882 | 0.03629 | 0.03317 | 0.02975 | 0.0263 | 0.023 | 0.01995 | 0.01723 | 0.01485 | 0.01279 | 0.01103 |
| 8 | 0.03388 | 0.03346 | 0.03224 | 0.03036 | 0.02802 | 0.02541 | 0.02273 | 0.02011 | 0.01765 | 0.01541 | 0.01342 | 0.01167 | 0.01014 |
| 8.5 | 0.02828 | 0.02797 | 0.02706 | 0.02565 | 0.02386 | 0.02184 | 0.01974 | 0.01764 | 0.01565 | 0.0138 | 0.01213 | 0.01064 | 0.00932 |
| 9 | 0.02385 | 0.02361 | 0.02292 | 0.02184 | 0.02047 | 0.01889 | 0.01722 | 0.01554 | 0.01391 | 0.01238 | 0.01097 | 0.0097 | 0.00857 |
| 9.5 | 0.0203 | 0.02012 | 0.01959 | 0.01875 | 0.01767 | 0.01643 | 0.0151 | 0.01374 | 0.0124 | 0.01113 | 0.00994 | 0.00886 | 0.00788 |
| 10 | 0.01742 | 0.01728 | 0.01686 | 0.01621 | 0.01536 | 0.01437 | 0.01329 | 0.01218 | 0.01108 | 0.01002 | 0.00902 | 0.00809 | 0.00724 |

**$B_{zz}/M_s^2 a^2 b$**

| $Z_0/a$ | 0 | 0.5 | 1.0 | 1.5 | 2.0 | 2.5 | 3.0 | 3.5 | 4.0 | 4.5 | 5.0 | 5.5 | 6.0 |
|---|---|---|---|---|---|---|---|---|---|---|---|---|---|
| $X_0/b$ | | | | | | | | | | | | | |
| 0 | | | | | -1.98974 | -0.85072 | -0.45312 | -0.27244 | -0.17735 | -0.1222 | -0.08789 | -0.06539 | -0.05 |
| 0.5 | | | | | -1.68964 | -0.78697 | -0.43213 | -0.26375 | -0.17319 | -0.11999 | -0.08663 | -0.06463 | -0.04951 |
| 1 | | | | | -1.12931 | -0.6332 | -0.37678 | -0.23983 | -0.16145 | -0.11367 | -0.08298 | -0.0624 | -0.04808 |
| 1.5 | | | | | -0.66535 | -0.45852 | -0.30419 | -0.20602 | -0.14412 | -0.10408 | -0.07733 | -0.05889 | -0.04582 |
| 2 | 1.43835 | 1.01724 | 0.22798 | -0.25837 | -0.3628 | -0.30803 | -0.23081 | -0.16847 | -0.1237 | -0.09232 | -0.07022 | -0.0544 | -0.04288 |
| 2.5 | 0.83664 | 0.64458 | 0.24903 | -0.05779 | -0.18364 | -0.19546 | -0.1667 | -0.13216 | -0.10259 | -0.07959 | -0.06225 | -0.04925 | -0.03944 |
| 3 | 0.52383 | 0.4282 | 0.21666 | 0.02452 | -0.08256 | -0.11769 | -0.11555 | -0.10012 | -0.08257 | -0.06689 | -0.054 | -0.04376 | -0.03569 |
| 3.5 | 0.34732 | 0.29615 | 0.17656 | 0.05478 | -0.0274 | -0.0665 | -0.07709 | -0.07355 | -0.06474 | -0.05496 | -0.04593 | -0.03821 | -0.03182 |
| 4 | 0.24106 | 0.21197 | 0.14104 | 0.06235 | 0.00166 | -0.03389 | -0.04932 | -0.05251 | -0.04958 | -0.04424 | -0.03838 | -0.03285 | -0.02797 |
| 4.5 | 0.17363 | 0.15624 | 0.11237 | 0.0604 | 0.01616 | -0.01365 | -0.02984 | -0.03636 | -0.03713 | -0.03496 | -0.03156 | -0.02784 | -0.02428 |
| 5 | 0.12895 | 0.11808 | 0.08996 | 0.05486 | 0.02263 | -0.00142 | -0.01648 | -0.02429 | -0.02718 | -0.02714 | -0.02555 | -0.02328 | -0.02083 |
| 5.5 | 0.09825 | 0.09121 | 0.07259 | 0.04838 | 0.02476 | 0.00572 | -0.00749 | -0.01543 | -0.01938 | -0.02068 | -0.0204 | -0.01924 | -0.01767 |
| 6 | 0.07651 | 0.0718 | 0.05912 | 0.04207 | 0.02463 | 0.00966 | -0.00156 | -0.00902 | -0.01337 | -0.01546 | -0.01604 | -0.0157 | -0.01484 |
| 6.5 | 0.06069 | 0.05745 | 0.0486 | 0.03637 | 0.02336 | 0.01161 | 0.00225 | -0.00447 | -0.00881 | -0.01128 | -0.01242 | -0.01265 | -0.01233 |
| 7 | 0.04893 | 0.04664 | 0.04032 | 0.0314 | 0.02159 | 0.01236 | 0.00462 | -0.00128 | -0.00538 | -0.00799 | -0.00945 | -0.01007 | -0.01014 |
| 7.5 | 0.04 | 0.03836 | 0.03376 | 0.02714 | 0.01966 | 0.01238 | 0.00601 | 9.24E-4 | -0.00284 | -0.00542 | -0.00703 | -0.00791 | -0.00825 |
| 8 | 0.03312 | 0.0319 | 0.0285 | 0.02352 | 0.01776 | 0.01198 | 0.00675 | 0.0024 | -9.82E-4 | -0.00343 | -0.00508 | -0.00611 | -0.00664 |
| 8.5 | 0.02772 | 0.02681 | 0.02424 | 0.02044 | 0.01596 | 0.01135 | 0.00706 | 0.00336 | 3.65E-4 | -0.00191 | -0.00353 | -0.00462 | -0.00528 |
| 9 | 0.02342 | 0.02273 | 0.02078 | 0.01784 | 0.01432 | 0.01062 | 0.00708 | 0.00394 | 0.00132 | -7.51E-4 | -0.0023 | -0.0034 | -0.00412 |
| 9.5 | 0.01997 | 0.01944 | 0.01792 | 0.01563 | 0.01284 | 0.00985 | 0.00693 | 0.00427 | 0.00198 | 1.17E-4 | -0.00133 | -0.00241 | -0.00316 |
| 10 | 0.01716 | 0.01675 | 0.01556 | 0.01375 | 0.01152 | 0.00909 | 0.00667 | 0.00441 | 0.00243 | 7.6E-4 | -5.77E-4 | -0.00161 | -0.00236 |

**$B_{xz}/M_s^2 a^2 b$**

| $Z_0/a$ | 0 | 0.5 | 1.0 | 1.5 | 2.0 | 2.5 | 3.0 | 3.5 | 4.0 | 4.5 | 5.0 | 5.5 | 6.0 |
|---|---|---|---|---|---|---|---|---|---|---|---|---|---|
| $X_0/b$ | | | | | | | | | | | | | |
| 0 | | | | | 0 | 0 | 0 | 0 | 0 | 0 | 0 | 0 | 0 |
| 0.5 | | | | | -0.58101 | -0.18863 | -0.08108 | -0.04099 | -0.02306 | -0.014 | -0.00901 | -0.00607 | -0.00424 |
| 1 | | | | | -0.79029 | -0.31015 | -0.14381 | -0.07551 | -0.04343 | -0.02675 | -0.01738 | -0.01178 | -0.00827 |
| 1.5 | | | | | -0.74145 | -0.35119 | -0.17946 | -0.09951 | -0.05916 | -0.03724 | -0.02456 | -0.01683 | -0.01191 |
| 2 | 0 | -1.07679 | -1.36032 | -1.02101 | -0.60607 | -0.33649 | -0.18997 | -0.11211 | -0.06941 | -0.04492 | -0.03021 | -0.02101 | -0.01503 |
| 2.5 | 0 | -0.55529 | -0.77415 | -0.67292 | -0.46705 | -0.29502 | -0.18274 | -0.11495 | -0.0744 | -0.04968 | -0.03419 | -0.02418 | -0.01753 |
| 3 | 0 | -0.31079 | -0.4659 | -0.4504 | -0.35123 | -0.24624 | -0.16561 | -0.11077 | -0.07499 | -0.05177 | -0.03653 | -0.02633 | -0.01937 |
| 3.5 | 0 | -0.18561 | -0.29377 | -0.30748 | -0.2622 | -0.20004 | -0.14451 | -0.10231 | -0.07237 | -0.05167 | -0.03742 | -0.02753 | -0.02059 |
| 4 | 0 | -0.11685 | -0.19273 | -0.21438 | -0.19606 | -0.16027 | -0.12312 | -0.09176 | -0.06766 | -0.04994 | -0.03713 | -0.0279 | -0.02122 |
| 4.5 | 0 | -0.07685 | -0.13085 | -0.15261 | -0.14756 | -0.12767 | -0.10339 | -0.08067 | -0.06182 | -0.0471 | -0.03594 | -0.02758 | -0.02134 |
| 5 | 0 | -0.05242 | -0.09152 | -0.11082 | -0.11207 | -0.10163 | -0.08609 | -0.06997 | -0.05555 | -0.04361 | -0.03411 | -0.02673 | -0.02105 |
| 5.5 | 0 | -0.03689 | -0.06569 | -0.08197 | -0.08599 | -0.08111 | -0.07141 | -0.06016 | -0.04931 | -0.03981 | -0.03189 | -0.02551 | -0.02043 |
| 6 | 0 | -0.02666 | -0.04825 | -0.06169 | -0.06669 | -0.06501 | -0.05918 | -0.05145 | -0.04341 | -0.03596 | -0.02947 | -0.02403 | -0.01957 |
| 6.5 | 0 | -0.01972 | -0.03615 | -0.04716 | -0.05228 | -0.05241 | -0.04909 | -0.04389 | -0.03801 | -0.03223 | -0.02698 | -0.02241 | -0.01855 |
| 7 | 0 | -0.01488 | -0.02757 | -0.03658 | -0.04141 | -0.04251 | -0.04081 | -0.03739 | -0.03315 | -0.02873 | -0.02452 | -0.02073 | -0.01743 |

| 7.5 | 0 | -0.01143 | -0.02137 | -0.02875 | -0.03313 | -0.03471 | -0.03405 | -0.03188 | -0.02886 | -0.02551 | -0.02217 | -0.01906 | -0.01627 |
| 8 | 0 | -0.00892 | -0.01681 | -0.02287 | -0.02676 | -0.02853 | -0.02851 | -0.02721 | -0.02511 | -0.02259 | -0.01997 | -0.01743 | -0.01509 |
| 8.5 | 0 | -0.00706 | -0.01338 | -0.01839 | -0.02181 | -0.0236 | -0.02399 | -0.02328 | -0.02184 | -0.01998 | -0.01793 | -0.01588 | -0.01393 |
| 9 | 0 | -0.00566 | -0.01078 | -0.01495 | -0.01792 | -0.01965 | -0.02026 | -0.01997 | -0.01902 | -0.01766 | -0.01608 | -0.01443 | -0.01281 |
| 9.5 | 0 | -0.00459 | -0.00878 | -0.01226 | -0.01484 | -0.01647 | -0.01719 | -0.01717 | -0.01659 | -0.01561 | -0.0144 | -0.01308 | -0.01175 |
| 10 | 0 | -0.00375 | -0.00722 | -0.01015 | -0.01239 | -0.01388 | -0.01466 | -0.01482 | -0.01449 | -0.0138 | -0.01289 | -0.01185 | -0.01076 |

**Table S3, a/b = 2.0**

$B_{xx}/M_s^2 a^2 b$

| $Z_0/a$ | 0 | 0.5 | 1.0 | 1.5 | 2.0 | 2.5 | 3.0 | 3.5 | 4.0 | 4.5 | 5.0 | 5.5 | 6.0 |
|---|---|---|---|---|---|---|---|---|---|---|---|---|---|
| $X_0/b$ | | | | | | | | | | | | | |
| 0 | | | | | 0.52961 | 0.19974 | 0.10203 | 0.0601 | 0.03866 | 0.02644 | 0.01892 | 0.01402 | 0.01069 |
| 0.5 | | | | | 0.42119 | 0.18422 | 0.09754 | 0.05835 | 0.03785 | 0.02602 | 0.01868 | 0.01388 | 0.0106 |
| 1 | | | | | 0.24104 | 0.14563 | 0.0853 | 0.0534 | 0.03552 | 0.02479 | 0.01798 | 0.01346 | 0.01034 |
| 1.5 | | | | | 0.10200 | 0.0995 | 0.06831 | 0.04608 | 0.03194 | 0.02288 | 0.01688 | 0.01279 | 0.00991 |
| 2 | -2.64183 | -2.09435 | -1.01502 | -0.25394 | 0.01406 | 0.05735 | 0.0499 | 0.03742 | 0.02751 | 0.02043 | 0.01544 | 0.0119 | 0.00933 |
| 2.5 | -1.54034 | -1.26785 | -0.69959 | -0.24471 | -0.03602 | 0.02406 | 0.03257 | 0.02844 | 0.02263 | 0.01763 | 0.01376 | 0.01084 | 0.00864 |
| 3 | -0.97314 | -0.82636 | -0.50525 | -0.21969 | -0.06159 | 1.506E-4 | 0.01774 | 0.01991 | 0.01769 | 0.01468 | 0.01193 | 0.00966 | 0.00786 |
| 3.5 | -0.65169 | -0.56773 | -0.37635 | -0.19070 | -0.0723 | -0.01577 | 0.00592 | 0.01234 | 0.01298 | 0.01172 | 0.01003 | 0.00842 | 0.00702 |
| 4 | -0.45644 | -0.40601 | -0.28696 | -0.16286 | -0.07445 | -0.02557 | -0.00298 | 0.00595 | 0.0087 | 0.0089 | 0.00816 | 0.00715 | 0.00614 |
| 4.5 | -0.33134 | -0.29978 | -0.22302 | -0.13807 | -0.07193 | -0.03098 | -0.00934 | 7.975E-4 | 0.00498 | 0.0063 | 0.00636 | 0.0059 | 0.00526 |
| 5 | -0.24769 | -0.22723 | -0.1762 | -0.11682 | -0.06710 | -0.03339 | -0.01363 | -0.0032 | 0.00183 | 0.00399 | 0.00469 | 0.0047 | 0.0044 |
| 5.5 | -0.18975 | -0.17607 | -0.14122 | -0.09892 | -0.06130 | -0.03383 | -0.01632 | -0.00618 | -7.346E-4 | 0.00197 | 0.00317 | 0.00357 | 0.00356 |
| 6 | -0.14842 | -0.13903 | -0.11466 | -0.08400 | -0.05531 | -0.03303 | -0.01780 | -0.00831 | -0.00277 | 2.721E-4 | 0.00183 | 0.00254 | 0.00278 |
| 6.5 | -0.11818 | -0.11158 | -0.09417 | -0.07160 | -0.04955 | -0.03150 | -0.01842 | -0.00974 | -0.00433 | -0.00113 | 6.646E-4 | 0.00161 | 0.00205 |
| 7 | -0.09557 | -0.09083 | -0.07817 | -0.06130 | -0.04421 | -0.02958 | -0.01844 | -0.01064 | -0.00549 | -0.00227 | -3.278E-4 | 7.8851E-4 | 0.00139 |
| 7.5 | -0.07835 | -0.07488 | -0.06550 | -0.05273 | -0.03938 | -0.02749 | -0.01804 | -0.01111 | -0.00632 | -0.00316 | -0.00115 | 7.3040E-5 | 7.941E-4 |
| 8 | -0.06500 | -0.06242 | -0.05537 | -0.04558 | -0.03506 | -0.02537 | -0.01737 | -0.01127 | -0.00687 | -0.00385 | -0.00183 | -5.400E-4 | 2.674E-4 |
| 8.5 | -0.05714 | -0.05255 | -0.04717 | -0.03958 | -0.03123 | -0.02330 | -0.01654 | -0.0112 | -0.00720 | -0.00435 | -0.00238 | -0.00106 | -1.928E-4 |
| 9 | -0.04814 | -0.04639 | -0.04049 | -0.03454 | -0.02785 | -0.02134 | -0.01562 | -0.01096 | -0.00736 | -0.00471 | -0.00280 | -0.00149 | -5.898E-4 |
| 9.5 | -0.04093 | -0.0396 | -0.03591 | -0.03066 | -0.02488 | -0.01952 | -0.01467 | -0.01062 | -0.00739 | -0.00494 | -0.00313 | -0.00184 | -9.276E-4 |
| 10 | -0.03509 | -0.03406 | -0.03118 | -0.02702 | -0.02228 | -0.01783 | -0.01372 | -0.0102 | -0.00731 | -0.00507 | -0.00336 | -0.00212 | -0.00121 |

$B_{yy}/M_s^2 a^2 b$

| $Z_0/a$ | 0 | 0.5 | 1.0 | 1.5 | 2.0 | 2.5 | 3.0 | 3.5 | 4.0 | 4.5 | 5.0 | 5.5 | 6.0 |
|---|---|---|---|---|---|---|---|---|---|---|---|---|---|
| $X_0/b$ | | | | | | | | | | | | | |
| 0 | | | | | 0.52961 | 0.19974 | 0.10203 | 0.0601 | 0.03866 | 0.02644 | 0.01892 | 0.01402 | 0.01069 |
| 0.5 | | | | | 0.49042 | 0.19446 | 0.10052 | 0.05952 | 0.03839 | 0.0263 | 0.01884 | 0.01397 | 0.01066 |
| 1 | | | | | 0.41016 | 0.18033 | 0.09624 | 0.05782 | 0.0376 | 0.02588 | 0.0186 | 0.01383 | 0.01057 |
| 1.5 | | | | | 0.32909 | 0.16109 | 0.08984 | 0.05518 | 0.03634 | 0.02522 | 0.01822 | 0.0136 | 0.01043 |
| 2 | 1.64798 | 1.42371 | 0.93869 | 0.50947 | 0.26038 | 0.14027 | 0.08216 | 0.05184 | 0.0347 | 0.02433 | 0.01771 | 0.01329 | 0.01023 |
| 2.5 | 0.91305 | 0.81203 | 0.58227 | 0.35729 | 0.2056 | 0.1202 | 0.07394 | 0.04805 | 0.03278 | 0.02328 | 0.0171 | 0.01291 | 0.00998 |
| 3 | 0.55615 | 0.50565 | 0.38599 | 0.25892 | 0.16294 | 0.10206 | 0.06577 | 0.04406 | 0.03067 | 0.02209 | 0.01639 | 0.01247 | 0.00969 |
| 3.5 | 0.36265 | 0.33536 | 0.26847 | 0.19261 | 0.12996 | 0.08628 | 0.05803 | 0.04006 | 0.02848 | 0.02082 | 0.01562 | 0.01198 | 0.00937 |
| 4 | 0.24901 | 0.23332 | 0.19377 | 0.14646 | 0.10448 | 0.07286 | 0.05093 | 0.03618 | 0.02627 | 0.01951 | 0.0148 | 0.01145 | 0.00902 |
| 4.5 | 0.17806 | 0.16857 | 0.1441 | 0.1135 | 0.08471 | 0.06158 | 0.04456 | 0.03252 | 0.0241 | 0.01818 | 0.01396 | 0.01091 | 0.00866 |
| 5 | 0.13157 | 0.12559 | 0.10985 | 0.08944 | 0.06927 | 0.05219 | 0.03894 | 0.02914 | 0.02203 | 0.01687 | 0.01312 | 0.01034 | 0.00827 |
| 5.5 | 0.09988 | 0.09597 | 0.08551 | 0.07152 | 0.05714 | 0.04438 | 0.03403 | 0.02606 | 0.02007 | 0.01561 | 0.01228 | 0.00978 | 0.00788 |
| 6 | 0.07756 | 0.07492 | 0.06777 | 0.05796 | 0.04752 | 0.03788 | 0.02977 | 0.02328 | 0.01825 | 0.0144 | 0.01146 | 0.00922 | 0.00749 |
| 6.5 | 0.06139 | 0.05957 | 0.05456 | 0.04753 | 0.03983 | 0.03248 | 0.02608 | 0.02079 | 0.01657 | 0.01326 | 0.01068 | 0.00867 | 0.0071 |
| 7 | 0.04941 | 0.04812 | 0.04453 | 0.0394 | 0.03348 | 0.02798 | 0.0229 | 0.01857 | 0.01503 | 0.01219 | 0.00993 | 0.00813 | 0.00671 |
| 7.5 | 0.04035 | 0.03941 | 0.03679 | 0.03298 | 0.02857 | 0.02396 | 0.02015 | 0.01661 | 0.01363 | 0.01119 | 0.00921 | 0.00762 | 0.00634 |
| 8 | 0.03336 | 0.03267 | 0.03072 | 0.02785 | 0.02452 | 0.0209 | 0.01755 | 0.01487 | 0.01237 | 0.01027 | 0.00854 | 0.00713 | 0.00598 |
| 8.5 | 0.0279 | 0.02738 | 0.02591 | 0.02372 | 0.02116 | 0.0183 | 0.01558 | 0.01333 | 0.01123 | 0.00943 | 0.00792 | 0.00666 | 0.00563 |
| 9 | 0.02356 | 0.02316 | 0.02204 | 0.02034 | 0.01837 | 0.01608 | 0.01386 | 0.01184 | 0.0102 | 0.00865 | 0.00733 | 0.00622 | 0.00529 |
| 9.5 | 0.02007 | 0.01977 | 0.0189 | 0.01757 | 0.01602 | 0.01418 | 0.01237 | 0.01068 | 0.00927 | 0.00794 | 0.00679 | 0.00581 | 0.00497 |
| 10 | 0.01755 | 0.01729 | 0.01654 | 0.01542 | 0.01404 | 0.01255 | 0.01106 | 0.00965 | 0.00835 | 0.00729 | 0.00629 | 0.00542 | 0.00467 |

**$B_{zz}/M_s^2 a^2 b$**

| $Z_0/a$<br>$X_0/b$ | 0 | 0.5 | 1.0 | 1.5 | 2.0 | 2.5 | 3.0 | 3.5 | 4.0 | 4.5 | 5.0 | 5.5 | 6.0 |
|---|---|---|---|---|---|---|---|---|---|---|---|---|---|
| 0 | | | | | -1.05903 | -0.39946 | -0.20406 | -0.12021 | -0.07732 | -0.05287 | -0.03783 | -0.02804 | -0.02138 |
| 0.5 | | | | | -0.91153 | -0.37866 | -0.19805 | -0.11787 | -0.07624 | -0.05231 | -0.03752 | -0.02785 | -0.02126 |
| 1 | | | | | -0.65117 | -0.32596 | -0.18153 | -0.11122 | -0.07311 | -0.05067 | -0.03659 | -0.02729 | -0.02091 |
| 1.5 | | | | | -0.43109 | -0.26059 | -0.15815 | -0.10125 | -0.06828 | -0.04809 | -0.0351 | -0.02639 | -0.02033 |
| 2 | 0.994 | 0.67076 | 0.07634 | -0.25553 | -0.27444 | -0.19762 | -0.13206 | -0.08925 | -0.0622 | -0.04476 | -0.03316 | -0.02519 | -0.01956 |
| 2.5 | 0.62736 | 0.45586 | 0.11733 | -0.11258 | -0.16958 | -0.14426 | -0.10651 | -0.07649 | -0.0554 | -0.04091 | -0.03086 | -0.02375 | -0.01862 |
| 3 | 0.41702 | 0.32073 | 0.11926 | -0.03924 | -0.10135 | -0.10221 | -0.08351 | -0.06398 | -0.04836 | -0.03677 | -0.02832 | -0.02213 | -0.01756 |
| 3.5 | 0.28906 | 0.23239 | 0.10789 | -0.00191 | -0.05766 | -0.07051 | -0.06395 | -0.0524 | -0.04145 | -0.03254 | -0.02565 | -0.0204 | -0.01639 |
| 4 | 0.20744 | 0.17269 | 0.09319 | 0.01639 | -0.03003 | -0.04729 | -0.04795 | -0.04213 | -0.03497 | -0.02841 | -0.02296 | -0.0186 | -0.01517 |
| 4.5 | 0.15329 | 0.13121 | 0.07892 | 0.02457 | -0.01278 | -0.0306 | -0.03522 | -0.03332 | -0.02908 | -0.02449 | -0.02032 | -0.0168 | -0.01392 |
| 5 | 0.11613 | 0.10165 | 0.06635 | 0.02738 | -0.00218 | -0.01879 | -0.02531 | -0.02594 | -0.02386 | -0.02086 | -0.01781 | -0.01504 | -0.01267 |
| 5.5 | 0.08988 | 0.08012 | 0.05572 | 0.0274 | 0.00417 | -0.01055 | -0.01771 | -0.01988 | -0.01934 | -0.01758 | -0.01545 | -0.01335 | -0.01145 |
| 6 | 0.07087 | 0.06412 | 0.04689 | 0.02604 | 0.00781 | -0.00485 | -0.01196 | -0.01497 | -0.01548 | -0.01467 | -0.0133 | -0.01176 | -0.01027 |
| 6.5 | 0.05679 | 0.05202 | 0.03962 | 0.02407 | 0.00174 | -9.775E-4 | -0.00765 | -0.01104 | -0.01223 | -0.01212 | -0.01134 | -0.01028 | -0.00915 |
| 7 | 0.04616 | 0.04272 | 0.03365 | 0.02191 | 0.00219 | 0.00161 | -0.00445 | -0.00794 | -0.00954 | -0.00992 | -0.0096 | -0.00892 | -0.0081 |
| 7.5 | 0.038 | 0.03548 | 0.02872 | 0.01976 | 0.0024 | 4.6E-4 | -0.00211 | -0.0055 | -0.00731 | -0.00803 | -0.00806 | -0.00769 | -0.00713 |
| 8 | 0.03164 | 0.02975 | 0.02465 | 0.01773 | 0.00245 | 8.225E-4 | -4.1E-4 | -0.0036 | -0.00549 | -0.00643 | -0.00671 | -0.00659 | -0.00624 |
| 8.5 | 0.02661 | 0.02518 | 0.02127 | 0.01587 | 0.00241 | 0.00105 | 8.99E-6 | -0.00214 | -0.00402 | -0.00507 | -0.00554 | -0.0056 | -0.00543 |
| 9 | 0.02258 | 0.02148 | 0.01846 | 0.0142 | 0.00232 | 0.00118 | 2.675E-4 | -0.00101 | -0.00283 | -0.00394 | -0.00453 | -0.00473 | -0.0047 |
| 9.5 | 0.01932 | 0.01847 | 0.0161 | 0.01271 | 0.00219 | 0.00124 | 4.475E-4 | -1.61E-4 | -0.00188 | -0.003 | -0.00366 | -0.00397 | -0.00405 |
| 10 | 0.01666 | 0.01598 | 0.01411 | 0.01139 | 0.00206 | 0.00126 | 5.7E-4 | 3.23E-5 | -0.00112 | -0.00222 | -0.00292 | -0.0033 | -0.00346 |

**B$_{xz}$/M$_s^2$a$^2$b**

| Z$_0$/a | 0 | 0.5 | 1.0 | 1.5 | 2.0 | 2.5 | 3.0 | 3.5 | 4.0 | 4.5 | 5.0 | 5.5 | 6.0 |
|---|---|---|---|---|---|---|---|---|---|---|---|---|---|
| X$_0$/b | | | | | | | | | | | | | |
| 0 | | | | | 0 | 0 | 0 | 0 | 0 | 0 | 0 | 0 | 0 |
| 0.5 | | | | | -0.28433 | -0.07368 | -0.02916 | -0.01415 | -0.00778 | -0.00466 | -0.00297 | -0.00198 | -0.00137 |
| 1 | | | | | -0.39213 | -0.12767 | -0.0539 | -0.0269 | -0.015 | -0.00906 | -0.00581 | -0.0039 | -0.00272 |
| 1.5 | | | | | -0.38853 | -0.15551 | -0.07148 | -0.03716 | -0.02121 | -0.01299 | -0.0084 | -0.00568 | -0.00398 |
| 2 | 0 | -0.82218 | -0.99905 | -0.67849 | -0.34182 | -0.16179 | -0.08136 | -0.04444 | -0.02611 | -0.0163 | -0.01068 | -0.00728 | -0.00513 |
| 2.5 | 0 | -0.46968 | -0.61708 | -0.47939 | -0.28498 | -0.15436 | -0.08461 | -0.0487 | -0.02959 | -0.01887 | -0.01256 | -0.00866 | -0.00616 |
| 3 | 0 | -0.28487 | -0.39896 | -0.34383 | -0.23141 | -0.13986 | -0.08296 | -0.05035 | -0.03169 | -0.02072 | -0.01403 | -0.0098 | -0.00703 |
| 3.5 | 0 | -0.18124 | -0.26744 | -0.25019 | -0.18557 | -0.12276 | -0.07814 | -0.04991 | -0.03257 | -0.02185 | -0.01508 | -0.01069 | -0.00775 |
| 4 | 0 | -0.11996 | -0.18477 | -0.18466 | -0.14809 | -0.10563 | -0.07159 | -0.04797 | -0.03245 | -0.02236 | -0.01573 | -0.01132 | -0.00831 |
| 4.5 | 0 | -0.0821 | -0.13101 | -0.13819 | -0.11815 | -0.0898 | -0.06433 | -0.04507 | -0.03156 | -0.02233 | -0.01604 | -0.01172 | -0.00872 |
| 5 | 0 | -0.05783 | -0.09502 | -0.10481 | -0.0945 | -0.07581 | -0.05703 | -0.04163 | -0.03013 | -0.02187 | -0.01604 | -0.01191 | -0.00897 |
| 5.5 | 0 | -0.04177 | -0.07033 | -0.0805 | -0.07591 | -0.06378 | -0.0501 | -0.03796 | -0.02834 | -0.0211 | -0.01579 | -0.01192 | -0.0091 |
| 6 | 0 | -0.03084 | -0.053 | -0.06259 | -0.06129 | -0.0536 | -0.04374 | -0.03429 | -0.02636 | -0.02011 | -0.01534 | -0.01178 | -0.00911 |
| 6.5 | 0 | -0.02322 | -0.04059 | -0.04922 | -0.04979 | -0.04507 | -0.03805 | -0.03077 | -0.02429 | -0.01896 | -0.01476 | -0.01151 | -0.00902 |
| 7 | 0 | -0.01778 | -0.03155 | -0.03913 | -0.04068 | -0.03796 | -0.03302 | -0.02747 | -0.02224 | -0.01774 | -0.01406 | -0.01114 | -0.00885 |
| 7.5 | 0 | -0.01383 | -0.02484 | -0.03141 | -0.03347 | -0.03206 | -0.02864 | -0.02443 | -0.02024 | -0.01648 | -0.0133 | -0.0107 | -0.0086 |
| 8 | 0 | -0.01091 | -0.01981 | -0.02546 | -0.0277 | -0.02715 | -0.02484 | -0.02169 | -0.01836 | -0.01524 | -0.0125 | -0.01021 | -0.00831 |
| 8.5 | 0 | -0.0087 | -0.01596 | -0.02082 | -0.02306 | -0.02309 | -0.02158 | -0.01923 | -0.0166 | -0.01402 | -0.0117 | -0.00968 | -0.00799 |
| 9 | 0 | -0.00703 | -0.01299 | -0.01716 | -0.01932 | -0.01969 | -0.01876 | -0.01704 | -0.01498 | -0.01287 | -0.0109 | -0.00914 | -0.00764 |
| 9.5 | 0 | -0.00573 | -0.01068 | -0.01426 | -0.01628 | -0.01686 | -0.01634 | -0.01511 | -0.0135 | -0.01178 | -0.01012 | -0.0086 | -0.00726 |
| 10 | 0 | -0.00473 | -0.00885 | -0.01193 | -0.01379 | -0.01449 | -0.01427 | -0.0134 | -0.01215 | -0.01077 | -0.00937 | -0.00806 | -0.00688 |

**Table S4, a/b = 3.0**

**B$_{xx}$/M$_s^2$a$^2$b**

| Z$_0$/a | 0 | 0.5 | 1.0 | 1.5 | 2.0 | 2.5 | 3.0 | 3.5 | 4.0 | 4.5 | 5.0 | 5.5 | 6.0 |
|---|---|---|---|---|---|---|---|---|---|---|---|---|---|
| X$_0$/b | | | | | | | | | | | | | |
| 0 | | | | | 0.19996 | 0.06455 | 0.03179 | 0.01842 | 0.01174 | 0.00798 | 0.00569 | 0.0042 | 0.0032 |
| 0.5 | | | | | 0.16427 | 0.06185 | 0.0311 | 0.01816 | 0.01162 | 0.00792 | 0.00566 | 0.00418 | 0.00319 |
| 1 | | | | | 0.11017 | 0.0546 | 0.02913 | 0.01742 | 0.01129 | 0.00775 | 0.00556 | 0.00413 | 0.00315 |
| 1.5 | | | | | 0.06639 | 0.04477 | 0.02616 | 0.01626 | 0.01075 | 0.00747 | 0.0054 | 0.00403 | 0.00309 |
| 2 | -1.83142 | -1.42122 | -0.63403 | -0.10576 | 0.03465 | 0.03427 | 0.02254 | 0.01476 | 0.01004 | 0.00709 | 0.00519 | 0.0039 | 0.00301 |
| 2.5 | -1.12155 | -0.89144 | -0.43213 | -0.09974 | 0.01259 | 0.02439 | 0.01863 | 0.01303 | 0.00919 | 0.00664 | 0.00493 | 0.00374 | 0.00291 |
| 3 | -0.74009 | -0.60172 | -0.31577 | -0.09307 | -0.00226 | 0.01579 | 0.01472 | 0.01119 | 0.00824 | 0.00612 | 0.00462 | 0.00356 | 0.00279 |
| 3.5 | -0.51453 | -0.42718 | -0.24089 | -0.08552 | -0.01194 | 0.00869 | 0.01104 | 0.00932 | 0.00725 | 0.00556 | 0.00429 | 0.00335 | 0.00265 |
| 4 | -0.372 | -0.31475 | -0.18913 | -0.07762 | -0.01795 | 0.00305 | 0.00771 | 0.0075 | 0.00623 | 0.00496 | 0.00393 | 0.00312 | 0.0025 |
| 4.5 | -0.27736 | -0.2387 | -0.15162 | -0.06985 | -0.02141 | -0.00128 | 0.00482 | 0.00578 | 0.00522 | 0.00436 | 0.00355 | 0.00288 | 0.00234 |
| 5 | -0.21206 | -0.18527 | -0.1235 | -0.0625 | -0.02313 | -0.00451 | 0.00236 | 0.00421 | 0.00426 | 0.00376 | 0.00317 | 0.00263 | 0.00217 |
| 5.5 | -0.16556 | -0.14659 | -0.10191 | -0.05573 | -0.02366 | -0.00683 | 3.296E-4 | 0.00281 | 0.00334 | 0.00317 | 0.00279 | 0.00237 | 0.002 |
| 6 | -0.13159 | -0.1179 | -0.085 | -0.04961 | -0.02342 | -0.00843 | -0.00131 | 0.00157 | 0.0025 | 0.00261 | 0.00241 | 0.00212 | 0.00182 |
| 6.5 | -0.10621 | -0.09616 | -0.07156 | -0.04413 | -0.02268 | -0.00947 | -0.0026 | 5.168E-4 | 0.00174 | 0.00208 | 0.00205 | 0.00187 | 0.00165 |
| 7 | -0.08689 | -0.07939 | -0.06075 | -0.03927 | -0.02164 | -0.01008 | -0.0036 | -3.772E-4 | 0.00105 | 0.00159 | 0.0017 | 0.00162 | 0.00147 |
| .5 | -0.07194 | -0.06627 | -0.05196 | -0.03499 | -0.02044 | -0.01036 | -0.00434 | -0.00112 | 4.525E-4 | 0.00114 | 0.00137 | 0.00138 | 0.0013 |
| 8 | -0.0602 | -0.05585 | -0.04474 | -0.03121 | -0.01917 | -0.01041 | -0.00488 | -0.00173 | -7.110E-5 | 7.262E-4 | 0.00106 | 0.00115 | 0.00113 |
| 8.5 | -0.05085 | -0.04748 | -0.03876 | -0.02789 | -0.01788 | -0.01028 | -0.00524 | -0.00221 | -5.204E-4 | 3.590E-4 | 7.757E-4 | 9.393E-4 | 9.696E-4 |
| 9 | -0.04332 | -0.04068 | -0.03377 | -0.02497 | -0.01662 | -0.01003 | -0.00547 | -0.00259 | -9.010E-4 | 3.329E-5 | 5.153E-4 | 7.374E-4 | 8.151E-4 |
| 9.5 | -0.0372 | -0.03511 | -0.02958 | -0.02241 | -0.01541 | -0.00969 | -0.00558 | -0.00288 | -0.00122 | -2.525E-4 | 2.789E-4 | 5.496E-4 | 6.686E-4 |
| 10 | -0.03216 | -0.03049 | -0.02604 | -0.02015 | -0.01426 | -0.00931 | -0.00561 | -0.00309 | -0.00148 | -5.009E-4 | 6.613E-5 | 3.761E-4 | 5.305E-4 |

**$B_{yy}/M_s^2 a^2 b$**

| $Z_0/a$ | 0 | 0.5 | 1.0 | 1.5 | 2.0 | 2.5 | 3.0 | 3.5 | 4.0 | 4.5 | 5.0 | 5.5 | 6.0 |
|---|---|---|---|---|---|---|---|---|---|---|---|---|---|
| $X_0/b$ | | | | | | | | | | | | | |
| 0 | | | | | 0.19996 | 0.06455 | 0.03179 | 0.01842 | 0.01174 | 0.00798 | 0.00569 | 0.0042 | 0.0032 |
| 0.5 | | | | | 0.18665 | 0.06364 | 0.03155 | 0.01833 | 0.0117 | 0.00796 | 0.00568 | 0.0042 | 0.0032 |
| 1 | | | | | 0.16171 | 0.06108 | 0.03088 | 0.01808 | 0.01159 | 0.0079 | 0.00564 | 0.00418 | 0.00318 |
| 1.5 | | | | | 0.13688 | 0.05733 | 0.02983 | 0.01768 | 0.0114 | 0.00781 | 0.00559 | 0.00415 | 0.00316 |
| 2 | 1.29489 | 1.08421 | 0.64484 | 0.28626 | 0.11506 | 0.05287 | 0.02847 | 0.01714 | 0.01115 | 0.00768 | 0.00552 | 0.0041 | 0.00314 |
| 2.5 | 0.74825 | 0.64188 | 0.41131 | 0.20846 | 0.09663 | 0.04814 | 0.02689 | 0.01649 | 0.01085 | 0.00752 | 0.00543 | 0.00405 | 0.0031 |
| 3 | 0.47136 | 0.41265 | 0.28121 | 0.15763 | 0.0813 | 0.04345 | 0.02519 | 0.01576 | 0.01049 | 0.00733 | 0.00532 | 0.00398 | 0.00306 |
| 3.5 | 0.31571 | 0.28117 | 0.20163 | 0.12235 | 0.06862 | 0.03897 | 0.02343 | 0.01498 | 0.0101 | 0.00712 | 0.0052 | 0.00391 | 0.00301 |
| 4 | 0.22147 | 0.20013 | 0.14973 | 0.09686 | 0.05815 | 0.03482 | 0.02167 | 0.01415 | 0.00968 | 0.00688 | 0.00506 | 0.00382 | 0.00296 |
| 4.5 | 0.16113 | 0.14741 | 0.11427 | 0.0779 | 0.04948 | 0.03104 | 0.01995 | 0.01332 | 0.00924 | 0.00664 | 0.00491 | 0.00373 | 0.0029 |
| 5 | 0.12073 | 0.11162 | 0.08915 | 0.0635 | 0.04229 | 0.02764 | 0.01831 | 0.01248 | 0.00879 | 0.00638 | 0.00476 | 0.00363 | 0.00283 |
| 5.5 | 0.09271 | 0.08648 | 0.07084 | 0.05236 | 0.03631 | 0.0246 | 0.01676 | 0.01167 | 0.00834 | 0.00611 | 0.0046 | 0.00353 | 0.00277 |
| 6 | 0.07267 | 0.06831 | 0.05718 | 0.04361 | 0.03132 | 0.02191 | 0.01532 | 0.01088 | 0.00788 | 0.00585 | 0.00443 | 0.00342 | 0.00269 |
| 6.5 | 0.05798 | 0.05486 | 0.04678 | 0.03665 | 0.02714 | 0.01953 | 0.01399 | 0.01012 | 0.00744 | 0.00558 | 0.00426 | 0.00331 | 0.00262 |
| 7 | 0.04697 | 0.04469 | 0.03872 | 0.03106 | 0.02361 | 0.01744 | 0.01277 | 0.0094 | 0.00701 | 0.00531 | 0.00409 | 0.0032 | 0.00254 |
| 7.5 | 0.03857 | 0.03688 | 0.0324 | 0.02652 | 0.02064 | 0.01559 | 0.01165 | 0.00872 | 0.00659 | 0.00505 | 0.00392 | 0.00309 | 0.00247 |
| 8 | 0.03204 | 0.03077 | 0.02736 | 0.02279 | 0.01811 | 0.01397 | 0.01064 | 0.00809 | 0.00619 | 0.00479 | 0.00375 | 0.00297 | 0.00239 |
| 8.5 | 0.0269 | 0.02593 | 0.0233 | 0.01972 | 0.01595 | 0.01254 | 0.00971 | 0.0075 | 0.00581 | 0.00454 | 0.00358 | 0.00286 | 0.00231 |
| 9 | 0.0228 | 0.02204 | 0.01999 | 0.01716 | 0.01411 | 0.01128 | 0.00887 | 0.00695 | 0.00545 | 0.0043 | 0.00342 | 0.00275 | 0.00223 |
| 9.5 | 0.01949 | 0.01889 | 0.01728 | 0.01501 | 0.01252 | 0.01016 | 0.00812 | 0.00644 | 0.0051 | 0.00406 | 0.00326 | 0.00264 | 0.00215 |
| 10 | 0.01678 | 0.01631 | 0.01502 | 0.01319 | 0.01116 | 0.00918 | 0.00743 | 0.00596 | 0.00478 | 0.00384 | 0.00311 | 0.00253 | 0.00208 |

**$B_{zz}/M_s^2 a^2 b$**

| $Z_0/a$ | 0 | 0.5 | 1.0 | 1.5 | 2.0 | 2.5 | 3.0 | 3.5 | 4.0 | 4.5 | 5.0 | 5.5 | 6.0 |
|---|---|---|---|---|---|---|---|---|---|---|---|---|---|
| $X_0/b$ | | | | | | | | | | | | | |
| 0 | | | | | -0.3998 | -0.1291 | -0.06357 | -0.03683 | -0.02347 | -0.01596 | -0.01138 | -0.00841 | -0.0064 |
| 0.5 | | | | | -0.35088 | -0.12548 | -0.06265 | -0.03649 | -0.02332 | -0.01588 | -0.01133 | -0.00838 | -0.00638 |
| 1 | | | | | -0.27186 | -0.11568 | -0.06001 | -0.0355 | -0.02287 | -0.01565 | -0.0112 | -0.00831 | -0.00633 |
| 1.5 | | | | | -0.20327 | -0.10209 | -0.05599 | -0.03393 | -0.02215 | -0.01528 | -0.01099 | -0.00818 | -0.00625 |

| $Z_0/a$ \ $X_0/b$ | 0 | 0.5 | 1.0 | 1.5 | 2.0 | 2.5 | 3.0 | 3.5 | 4.0 | 4.5 | 5.0 | 5.5 | 6.0 |
|---|---|---|---|---|---|---|---|---|---|---|---|---|---|
| 2 | 0.53672 | 0.33709 | -0.0108 | -0.1805 | -0.14971 | -0.08714 | -0.05101 | -0.0319 | -0.02119 | -0.01477 | -0.01071 | -0.00801 | -0.00614 |
| 2.5 | 0.37336 | 0.24959 | 0.02081 | -0.10872 | -0.10923 | -0.07253 | -0.04552 | -0.02953 | -0.02003 | -0.01416 | -0.01035 | -0.00779 | -0.00601 |
| 3 | 0.26875 | 0.18909 | 0.03456 | -0.06456 | -0.07904 | -0.05924 | -0.03991 | -0.02695 | -0.01874 | -0.01345 | -0.00994 | -0.00754 | -0.00585 |
| 3.5 | 0.19884 | 0.14601 | 0.03926 | -0.03683 | -0.05669 | -0.04766 | -0.03446 | -0.02429 | -0.01735 | -0.01267 | -0.00948 | -0.00725 | -0.00566 |
| 4 | 0.15054 | 0.11462 | 0.0394 | -0.01923 | -0.04019 | -0.03787 | -0.02938 | -0.02165 | -0.01591 | -0.01185 | -0.00899 | -0.00694 | -0.00546 |
| 4.5 | 0.11624 | 0.09129 | 0.03735 | -0.00806 | -0.02806 | -0.02975 | -0.02477 | -0.0191 | -0.01446 | -0.011 | -0.00846 | -0.00661 | -0.00524 |
| 5 | 0.09133 | 0.07366 | 0.03435 | -0.00101 | -0.01916 | -0.02313 | -0.02067 | -0.01669 | -0.01304 | -0.01014 | -0.00793 | -0.00626 | -0.00501 |
| 5.5 | 0.07286 | 0.06012 | 0.03106 | 0.00337 | -0.01265 | -0.01777 | -0.01709 | -0.01447 | -0.01168 | -0.00929 | -0.00738 | -0.0059 | -0.00477 |
| 6 | 0.05892 | 0.04959 | 0.02782 | 0.00599 | -0.0079 | -0.01348 | -0.01401 | -0.01245 | -0.01039 | -0.00846 | -0.00684 | -0.00554 | -0.00452 |
| 6.5 | 0.04823 | 0.0413 | 0.02479 | 0.00748 | -0.00446 | -0.01007 | -0.01139 | -0.01064 | -0.00918 | -0.00766 | -0.0063 | -0.00518 | -0.00427 |
| 7 | 0.03992 | 0.0347 | 0.02203 | 0.00821 | -0.00197 | -0.00736 | -0.00917 | -0.00902 | -0.00806 | -0.0069 | -0.00579 | -0.00482 | -0.00402 |
| 7.5 | 0.03337 | 0.02939 | 0.01956 | 0.00847 | -1.93E-4 | -0.00523 | -0.00731 | -0.0076 | -0.00704 | -0.00618 | -0.00529 | -0.00447 | -0.00377 |
| 8 | 0.02815 | 0.02508 | 0.01738 | 0.00842 | 0.00106 | -0.00356 | -0.00576 | -0.00636 | -0.00612 | -0.00551 | -0.00481 | -0.00413 | -0.00352 |
| 8.5 | 0.02395 | 0.02155 | 0.01546 | 0.00817 | 0.00193 | -0.00226 | -0.00447 | -0.00529 | -0.00529 | -0.0049 | -0.00436 | -0.0038 | -0.00328 |
| 9 | 0.02052 | 0.01864 | 0.01378 | 0.00782 | 0.00251 | -0.00125 | -0.00341 | -0.00436 | -0.00454 | -0.00433 | -0.00393 | -0.00349 | -0.00305 |
| 9.5 | 0.01771 | 0.01621 | 0.01231 | 0.0074 | 0.00288 | -4.7E-4 | -0.00253 | -0.00356 | -0.00388 | -0.00381 | -0.00354 | -0.00319 | -0.00282 |
| 10 | 0.01538 | 0.01418 | 0.01101 | 0.00696 | 0.00311 | 1.24E-4 | -0.00182 | -0.00287 | -0.0033 | -0.00334 | -0.00317 | -0.0029 | -0.00261 |

$B_{xz}/M_s^2 a^2 b$

| $Z_0/a$ \ $X_0/b$ | 0 | 0.5 | 1.0 | 1.5 | 2.0 | 2.5 | 3.0 | 3.5 | 4.0 | 4.5 | 5.0 | 5.5 | 6.0 |
|---|---|---|---|---|---|---|---|---|---|---|---|---|---|
| 0 | | | | | 0 | 0 | 0 | 0 | 0 | 0 | 0 | 0 | 0 |
| 0.5 | | | | | -0.09329 | -0.01747 | -0.00638 | -0.00299 | -0.00161 | -9.53E-4 | -6.03E-4 | -4E-4 | -2.77E-4 |
| 1 | | | | | -0.1314 | -0.03216 | -0.01226 | -0.00583 | -0.00317 | -0.00188 | -0.00119 | -7.96E-4 | -5.5E-4 |
| 1.5 | | | | | -0.13898 | -0.04259 | -0.01724 | -0.00841 | -0.00462 | -0.00277 | -0.00176 | -0.00118 | -8.16E-4 |
| 2 | 0 | -0.5089 | -0.58582 | -0.34741 | -0.13279 | -0.04869 | -0.02111 | -0.01061 | -0.00593 | -0.00358 | -0.0023 | -0.00154 | -0.00107 |
| 2.5 | 0 | -0.32534 | -0.39604 | -0.26277 | -0.12088 | -0.05119 | -0.02381 | -0.01239 | -0.00706 | -0.00432 | -0.00279 | -0.00188 | -0.00132 |
| 3 | 0 | -0.21748 | -0.27832 | -0.20281 | -0.10716 | -0.05105 | -0.02541 | -0.01373 | -0.008 | -0.00496 | -0.00323 | -0.0022 | -0.00154 |
| 3.5 | 0 | -0.15041 | -0.20125 | -0.15873 | -0.09352 | -0.04914 | -0.02608 | -0.01464 | -0.00873 | -0.0055 | -0.00362 | -0.00248 | -0.00175 |
| 4 | 0 | -0.10693 | -0.14883 | -0.12559 | -0.08087 | -0.04618 | -0.02601 | -0.01517 | -0.00927 | -0.00594 | -0.00396 | -0.00273 | -0.00194 |
| 4.5 | 0 | -0.0778 | -0.11214 | -0.10029 | -0.06955 | -0.04265 | -0.02537 | -0.01536 | -0.00963 | -0.00627 | -0.00423 | -0.00295 | -0.00211 |
| 5 | 0 | -0.05775 | -0.08586 | -0.08075 | -0.05964 | -0.03891 | -0.02432 | -0.01527 | -0.00982 | -0.00651 | -0.00445 | -0.00313 | -0.00226 |
| 5.5 | 0 | -0.04363 | -0.06668 | -0.06551 | -0.05109 | -0.03519 | -0.02302 | -0.01496 | -0.00987 | -0.00667 | -0.00462 | -0.00328 | -0.00239 |
| 6 | 0 | -0.03348 | -0.05243 | -0.05353 | -0.04377 | -0.03162 | -0.02157 | -0.01448 | -0.00979 | -0.00674 | -0.00473 | -0.0034 | -0.00249 |
| 6.5 | 0 | -0.02606 | -0.0417 | -0.04404 | -0.03753 | -0.0283 | -0.02005 | -0.01388 | -0.00961 | -0.00673 | -0.0048 | -0.00348 | -0.00258 |
| 7 | 0 | -0.02055 | -0.03351 | -0.03646 | -0.03224 | -0.02524 | -0.01852 | -0.0132 | -0.00935 | -0.00667 | -0.00482 | -0.00354 | -0.00264 |
| 7.5 | 0 | -0.01639 | -0.02719 | -0.03038 | -0.02774 | -0.02248 | -0.01702 | -0.01247 | -0.00903 | -0.00656 | -0.00481 | -0.00357 | -0.00268 |
| 8 | 0 | -0.01321 | -0.02225 | -0.02546 | -0.02393 | -0.02 | -0.01559 | -0.01171 | -0.00867 | -0.00641 | -0.00476 | -0.00357 | -0.00271 |
| 8.5 | 0 | -0.01075 | -0.01836 | -0.02145 | -0.0207 | -0.01778 | -0.01424 | -0.01096 | -0.00828 | -0.00622 | -0.00468 | -0.00355 | -0.00272 |
| 9 | 0 | -0.00883 | -0.01526 | -0.01818 | -0.01795 | -0.01582 | -0.01298 | -0.01021 | -0.00787 | -0.00601 | -0.00459 | -0.00352 | -0.00272 |
| 9.5 | 0 | -0.00731 | -0.01278 | -0.01548 | -0.01561 | -0.01408 | -0.01181 | -0.00949 | -0.00745 | -0.00578 | -0.00447 | -0.00347 | -0.0027 |
| 10 | 0 | -0.0061 | -0.01077 | -0.01325 | -0.01362 | -0.01254 | -0.01074 | -0.0088 | -0.00703 | -0.00553 | -0.00433 | -0.0034 | -0.00268 |